# Monographie sur le tolérancement modal

# 1  La genèse de l'idée

Le défaut de forme est un parent pauvre du tolérancement. Il est entre la rugosité pour laquelle des méthodes d'analyses issues du traitement du signal donnent des résultats intéressants (avec maintenant une ouverture au 3D) et les dimensions et positions des surfaces où les méthodes issues de la théorie des mécanismes s'appliquent. Ce n'est pas la volonté d'explorer ce thème qui a fait germer l'idée du tolérancement modal mais une rencontre avec un industriel facteur de cloches. La mise en place d'un TP en simulation et expérimentations a donné une coopération recherche-transfert [SAM 03a] et posé la question de la corrélation entre vibration sonore et défauts géométriques. Le problème du paramétrage s'est posé. L'idée du paramétrage modal était née.

## 1.1  Contexte et intérêt de la méthode

La maîtrise de la qualité des géométries se décompose en 6 ordres de défauts. Ces ordres sont définis selon des critères technologiques et mathématiques. En effet, d'un point de vue purement mathématique, si l'on s'intéresse aux expressions des variations d'une géométrie, on doit considérer le paramétrage qui nous permet de la définir. En effet, les variations possibles (imaginables ou observables) sont indépendantes du paramétrage même si ce dernier nous est indispensable pour l'appréhender.

**Nos constats :**
-        Il est nécessaire de définir des modèles de variations de formes géométriques pour tous ou chacun des ordres de défauts. Ces modèles définissent des langages permettant à la communauté d'échanger sur les limites (et leur écriture) admissibles par la géométrie observable par rapport à l'idéale.

-        Le matériel de métrologie a progressé en performances et surtout en nombre de points mesurés plus vite que les méthodes d'expressions de défauts.

-        Les besoins des clients en qualité géométriques en exigence quantitative mais aussi qualitative augmentent. En effet, il est de plus en plus courant de voir exprimés des besoins en qualité perçue sur nombre de pièces. Comment écrire ces besoins ? Les annotations autorisent à mettre noir sur blanc un besoin mais on comprend la difficulté à en assurer la qualité qui reste pour trop d'entre elles soumises à des appréciations et mettent le métrologue dans l'embarras.

-        Les pièces de petites dimensions (micro, voire nano mécanique) peuvent avoir des défauts géométriques dont les dimensions ne permettent plus la distinction en ordres décroissants. En effet, un petit diamètre de trou aura des défauts de formes qui pourront être bien plus importants que le défaut dimensionnel. Comment alors utiliser les méthodes traditionnelles pour exprimer le besoin ? L'exigence de l'enveloppe et l'utilisation du « maxi-matière » permettent de lier dimension-position et forme mais il peut être très utile aux acteurs du tolérancement d'utiliser un langage permettant de distinguer les différents défauts.

**Des réponses :**
-       Dans le domaine de l'optique, la forme est la fonction, il s'agit logiquement du domaine qui, le premier, a défini de façon rigoureuse une méthode d'écriture des défauts de forme à l'aide des polynômes de Zernike [WYA 92][ISO 10110-5]. Ce langage montre la voie à suivre pour l'écriture sur des géométries plus complexes que les disques. Il donne une expression de complexité croissante des formes et est associé (pour les premières formes) aux problèmes qu'on veut maîtriser.
-       Dans le domaine médical l'imagerie 2D et 3D crée des données très volumineuses à traitement quasi exclusivement humain. Cootes [COO 05] effectue une synthèse bibliographique (de 1994 à 2004) des méthodes de correspondance entre images et formes dans l'objectif de reconnaître (voire de paramétrer) les organes internes des patients. Dans ce domaine on trouve une très grande richesse de solutions en 2D et 3D. Les travaux de Nastar et Ayache [NAS 96] permettent de coder les organes à l'aide d'un paramétrage basé sur les modes naturels de leurs surfaces enveloppes.
-       Dans le domaine mécanique, le paramétrage des défauts de forme a été le plus souvent traité à l'aide de modèles utilisant les concepts des séries de Fourier [PAN 81] [HUA 02] [MOR 07] [CAP 00] [CAP 01a] [HEN 99] ou étendus à une optimisation statistique [CAP 01b] mais limités à des surfaces bi-paramétrées (plans, cylindres en général). Des descriptions à l'aide de fonctions polynomiales [PFE 01] sont limitées dans la complexité des défauts à identifier. Le paramétrage des surfaces déclarées dans une application de CAO telles les surfaces de Bézier peut être lui aussi utilisé [CUB 98].
-       Le procédé de mise en œuvre (enlèvement de matière en principe) peut être lui aussi la base d'un paramétrage [DUC 99][TOU 02][LAC 04] dans lequel la forme, l'ondulation voire la rugosité sont prises en compte.

# 1.2  Les six Ordres de défauts

Le tolérancement géométrique est l'expression des variations admissibles de la géométrie réalisée autour de l'idéal défini par le concepteur. Ces variations sont découpées ([ISO 8785] en 6 catégories (en principe ordonnées selon la valeur du défaut). Nous proposons de décomposer en trois types ces catégories, le premier est lié un modificateur paramétrique donné par l'intention du concepteur, la dimension, la seconde à la mobilité de la géométrie (position et orientation) enfin la dernière concerne les écarts autour de ces deux premières nous l'appelerons FOR pour Forme-Ondulation-Rugosité qui constitue un groupe cohérent. Nous nous intéresserons particulèrement à ce dernier. Néanmoins, nous donnons les définitions des 6 catégories ci-dessous :

1. **Dimension-Taille**
   Il s'agit en principe du paramètre géométrique (distance, diamètre, …) que l'on fait varier. Les points mesurés sont remplacés par une surface parfaite selon un critère d'association (moindres carrés, …).
2. **Position (Translation et Orientation)**
   Ici, il est nécessaire de définir un repère lié à la surface associée à la géométrie actuelle.

Les quatre catégories suivantes sont séparées par une opération de filtrage. Il s'agit des défauts de forme, d'ondulation et de rugosité. Mathématiquement, ils sont du même type mais pas de mêmes grandeurs. On peut les qualifier et les quantifier simplement en les observant selon une ligne. On pourra alors extraire les longueurs d'ondes et les amplitudes associées en utilisant une décomposition en séries de Fourier.

3. **Forme**

Les défauts sont de grandes longueurs d'ondes. Notre proposition est de définir la forme comme étant le défaut qui influe majoritairement sur la mise en position de la surface. Elle peut se réduire à un ensemble de points de contact iso-statiques. Les déformations potentielles issues d'une mise en application d'une force de liaison déformeront la surface complètement (étude de structure)

4. **Ondulation**

L'ondulation est en principe extraite de la rugosité la norme [ISO 4288] permet de définir le filtre de coupure.

Notre proposition est de la définir comme un défaut de longueur d'onde moyenne dont les nombreux points de contact potentiels sont des sous-surfaces lissées. La mise en contact de deux surfaces avec des défauts d'ondulation en est peu perturbée et les écrasements sont plutôt locaux (déformations du type Hertziens).

5. **Rugosité**

La géométrie est supposée périodique mais de très courte longueur d'onde. A ce jour les normes ne permettent que le traitement des profils mais une nouvelle norme permettra d'étudier les caractéristiques locales 3D [ISO 25178].

Notre proposition est de définir la rugosité comme un défaut de forte ondulation dont la mise en position de deux surfaces définira des micro-spots très peu influents sur la position et dont le comportement (quelquefois Hertziens) suite à une mise en charge pourra entrainer une plastification rapide.

6. **Micro rugosité**

Il n'y a pas à ce niveau de périodicité évidente. C'est pourquoi les outils d'analyses sont exclusivement statistiques.

Il est nécessaire de définir des modèles associés à chaque catégorie de façon à pouvoir écrire des spécifications claires, faire des simulations et effectuer des contrôles. Ces 6 ordres définissent la première étape de la modélisation géométrique. L'étape suivante est la définition d'un paramétrage. Les domaines de prédilection de nos travaux sont la forme et l'ondulation, même si, en théorie le paramétrage que nous introduisons permet de toutes les étudier.

## 1.3 Paramétrage

Les modèles de tolérancement s'appuient sur un paramétrage de la géométrie et de ses instances. Il doit permettre de décrire au mieux le réel observé ou (et) modélisé. La mesure des objets a permis de développer des modèles mathématiques de complexité croissante pour les analyser. Actuellement, les moyens de mesures permettent avec une facilité croissante et des coûts décroissants d'obtenir une mesure d'une géométrie par un grand nombre de points. Le traitement de ces données permet à de nouveaux modèles d'exprimer les objets mesurés avec une plus grande précision. Notre modèle ([SAM 07a]) décrit le défaut comme un ensemble de paramètres positionnels.

## 1.4  Base de défauts de formes

Une base est nécessaire pour pouvoir définir la nature du défaut, la spécifier, la contrôler, la piloter. Dans le langage courant, nous parlons de conicité d'un cylindre, d'une forme en tonneau, en diabolo, d'ovalisation et dans le cas d'un plan, nous parlerons d'un bombé, d'un vrillage (listes non exhaustives). Au-delà du vocabulaire technique, il est indispensable de définir une description mathématique des variations possibles d'une géométrie au voisinage de l'idéal qui la définit. Pour ce faire, nous avons les trois possibilités suivantes :

- Soit nous avons une méthode permettant cette description à priori des variations (défauts potentiels ou observés) et nous les définissons comme géométries particulières que nous appelons **défauts technologiques**,
- soit nous utilisons une famille d'équations analytiques permettant cette description que nous appelons **paramétrage explicite (ou intrinsèque)** qui s'applique à des géométries particulières,
- soit nous utilisons une famille d'équations discrètes qui permet de décrire tout type de variation de forme de toute géométrie que nous appelons **base naturelle à paramétrage implicite**.

Il nous a semblé intéressant de présenter ces trois types de base de défauts car elles sont au cœur des choix que font les technologues (et de normes) lors de la description des spécifications géométriques. Nous ne tranchons pas sur la pertinence du choix de l'une des trois solutions, et le cas échéant, il sera judicieux de les utiliser en complémentarité (cas des bases naturelles avec introduction des défauts technologiques).

**Illustrons notre propos avec un cylindre :**
Le plus naturel est de donner pour un cylindre, deux paramètres dimensionnels et quatre paramètres positionnels. On peut décrire sa conicité, en rajoutant un paramètre dimensionnel (on peut différentier les deux rayons extrêmes, par exemple), son bombé (ou diabolo) en ajoutant un autre paramètre dimensionnel comme un rayon intermédiaire (base technologique). On peut aussi utiliser l'équation du paraboloïde de révolution positionné dans un repère et utiliser ses paramètres (base explicite). On est alors confronté à la nécessité de trouver les formes que peut prendre la famille de cylindres observables. On peut, en utilisant des méthodes mathématiques, définir des bases analytiques exhaustives à l'instar de Gouskov [GOU 99] ou Summerhays [SUM 02]. On peut encore proposer une méthode qui définisse de façon systématique la base associée à la géométrie selon une théorie vibratoire, c'est ce que nous ferons par la suite. Nous appelons cette dernière base naturelle pour reprendre le terme « modes naturels ». Ce type de paramétrage sera appelé implicite car il est défini automatiquement dès que la géométrie idéale l'est, et ce sans aucune requête à l'utilisateur ou à un système expert.

# 2   Principe de la méthode modale

La base de défauts géométriques est construite à l'aide des modes naturels associés à la géométrie idéale. Nous discuterons des modélisations (géométries, conditions aux limites) à partir du §2.1. Toute géométrie, mesurée ou simulée peut être vue comme une géométrie déformée suivant un chargement. Cette modélisation d'un défaut nous amène à deux points de vue, le point de vue explicite [CAM 03a] dans lequel les formes sont construites à priori par des chargements d'une structure élastique et le point de vue implicite où les formes sont construites automatiquement. Dans un premier temps nous avons pensé que le second point de vue, plus général, serait suffisant pour l'utilisateur. En appliquant nos modèles à des cas industriels [FAV 07],[ADR 07b],[ADR 06a],[FOR 07] nous avons mis en œuvre des modèles mixant les deux points de vue.

Nous proposons une démarche de type « superposition modale » en utilisant la structure de base vectorielle des modes naturels pour déterminer la projection d'un champ de déplacements.

[MOK 92] définit six critères de choix d'un paramétrage dans le cas d'une base de formes topographiques :

1  **L'unicité** = un seul vecteur de paramètres représente la forme
2  **L'inversibilité** = bijection entre les deux espaces de représentations (cartésien et modal)
3  **La stabilité** = continuité du paramétrage
4  **L'invariance** = découplage des paramètres de position de taille et de forme
5  **L'accès aux propriétés géométriques** = détection aisée des propriétés géométriques
6  **Efficience et complexité algorithmique** = capacité du paramétrage à représenter le mieux possible la forme en limitant le nombre de paramètres

L'invariance (point 4) n'est pas forcément respectée par les formes modales, mais elle peut l'être par l'utilisation de modes technologiques. Notre paramétrage respecte ces six critères auxquels nous ajoutons les trois critères suivants :
7  **La complexité croissante des modes**
8  **L'exhaustivité** des formes variationnelles associées à la forme de référence.
9  **Une métrique** permettant d'associer à chaque paramètre modal une signification géométrique dimensionnelle.

## 2.1  Eléments de dynamique des milieux continus

Nous présentons dans ce paragraphe la mise en équation dynamique d'un milieu continu élastique en régime libre afin d'en extraire les modes naturels.

Tout point du domaine est défini par trois coordonnées et une variable de temps. Soient deux points distants de $\vec{dx}$, le champ de déplacement $\vec{U}$ se décompose en une translation $\vec{T}$, et un accroissement $\overline{\overline{\nabla}}(\vec{U}).\vec{dx}$ décomposé lui-même en une rotation $[\Omega].\vec{dx}$ et une déformation $[\varepsilon].\vec{dx}$

$$d\vec{U} = \vec{T} + \overline{\overline{\nabla}}(\vec{U}) \, \vec{dx} = \vec{T} + [\varepsilon].\vec{dx} + [\Omega].\vec{dx} \qquad (2.1.1)$$

On utilise les relations élastiques contraintes déformations (loi de Hooke) :
$$[\sigma] = 2\mu \, [\varepsilon] + \lambda \, \mathrm{tr}[\varepsilon] \; \mathrm{Id} \qquad (2.1.2)$$
Où $\lambda$ et $\mu$ sont les deux constantes de Lamé du matériau et $[\sigma]$ le tenseur des contraintes.

Les équations d'équilibre dynamique appliquées à un élément volumique infinitésimal dv, associées aux conditions aux limites s'écrivent :

$$\begin{cases} \rho \; \vec{\Gamma} = \vec{f_v} + \overrightarrow{\mathrm{div}([\sigma])} & \text{en tout point du volume} \\ [\sigma]\{\vec{n}\} = \vec{f_n} \; dS & \text{en tout point de la surface du domaine} \end{cases} \qquad (2.1.3)$$

La résolution des équations (2.1.1), (2.1.2) et (2.1.3) dans un domaine continu permet d'obtenir les solutions en déplacements en tout point du solide. Néanmoins, ces équations n'ont de solutions exploitables que dans des cas où la géométrie est simple, les poutres, les plaques à formes simples [LEI 69], et les membranes.

Afin de pouvoir utiliser les propriétés géométriques des modes naturels pour toute géométrie, nous avons opté pour la méthode des éléments finis. La figure 2.1.1 illustre la méthode en présentant les vingt premiers modes naturels de flexion d'un disque de faible épaisseur. Ils sont triés selon un critère fréquentiel. Nous ne nous intéressons pas à la valeur des fréquences mais seulement au tri qu'elles effectuent sur les modes. A ces formes sont affecté autant de paramètres de la géométrie que nous pouvons présenter à la manière d'un spectre (cf. figure 2.2.1 page 11). Les décompositions géométriques peuvent être vues comme celles des séries (formes harmoniques) de Fourier (dont les spectres sont bien connus du métrologue).

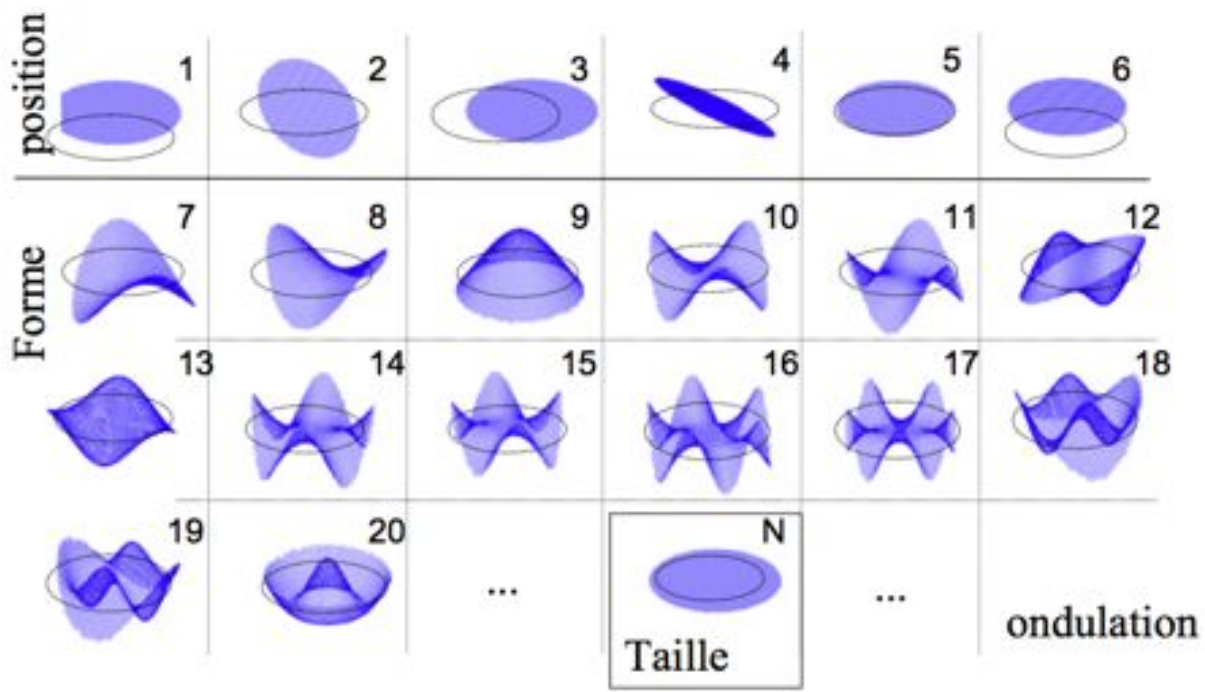

*Figure 2.1.1 Modes propres d'un disque*

Nous avons donc deux objectifs : celui de définir la base de défauts associée à la géométrie à étudier et celui d'effectuer la décomposition d'une géométrie actuelle dans cette base afin d'en déduire les composantes.

Comme le disque de la figure 2.1.1 n'a pas été contraint cinématiquement, les six premiers modes seront des modes rigides (qui sont des combinaisons des six mobilités). Ces modes peuvent être traduits en TPD de la surface associée. Les modes qui suivent sont qualifiés de modes de forme car leurs déformées sont de grandes périodes. Pour des valeurs plus élevées du mode, on trouvera des modes d'ondulation et un mode de taille à la position N qui variera selon le rapport épaisseur sur rayon.

Ces formes modales peuvent être obtenues à l'aide de modèles continus (équations) ou discrets (MEF). Dans le cas d'un modèle MEF, les formes sont définies par des champs de déplacements des points du maillage (nœuds) complétées par les fonctions d'interpolations du modèle d'élément utilisé. Les formes sont donc des données discrètes (déplacements nodaux) complétées par des fonctions continues (interpolations). Dans un premier temps, nous avons choisi de ne pas exploiter l'usage de ces fonctions d'interpolations, mais elles pourront êtres utiles pour effectuer des correspondances entre mesures et MEF.

## 2.2  Eléments de dynamique des systèmes discrets

La première opération est la discrétisation de la géométrie nominale. Dès lors le paramétrage s'appuie sur les fonctions d'interpolations associées aux éléments.

Nous transformons alors une géométrie en une structure mécanique discrète. Il est alors calculé [GER 96] une matrice de raideur K et une matrice de masse M. La détermination des modes naturels s'effectue en résolvant l'équation matricielle d'équilibre dynamique en petits mouvements (2.2.1).

$$M \ddot{q} + K q = 0 \qquad\qquad (2.2.1)$$

Les solutions (modes naturels) sont de la forme :

$$q_i = Q_i \cos(\omega_i t) \qquad\qquad (2.2.2)$$

Où $Q_i$ est un vecteur d'amplitudes et $\omega_i$ la pulsation associée.

Le calcul des modes naturels ($\omega_i, Q_i$) se fait en résolvant les équations (2.2.3)

$$\left( M^{-1} K - \frac{1}{\omega_i^2} I_d \right) q_i = 0 \qquad\qquad (2.2.3)$$

Les modes sont alors triés selon les $\omega_i$ croissants de façon à obtenir les déformées les « plus simples » en premier.

Un défaut est une surface (discrète ici) représentée par un vecteur de déplacements V (ensemble d'écarts géométriques relatifs au nominal). Ce vecteur se décompose dans la base modale ($Q_i$) afin d'obtenir l'ensemble de contributions $\lambda_i$ (coordonnées modale) de chaque déformée modale. Nous avons dans nos travaux, utilisé plusieurs types de projections. La confrontation aux besoins industriels nous a fait converger vers la méthode présentée dans les équations (2.2.4) (2.2.5) et (2.2.6).

Nous définissons la norme des $Q_i$ par :

$$\|Q_i\|_\infty = 1 \qquad\qquad (2.2.4)$$

Le vecteur V à projeter dans la base ($Q_1, Q_2$) s'écrit :

$$V = \sum_{i=1}^{n} \lambda_i Q_i = Q \lambda \qquad\qquad (2.2.5)$$

Où Q est la matrice des modes propres et $\lambda$ le vecteur des coefficients modaux. On obtient $\lambda$ en résolvant les équations (2.2.6).

$$(Q^t Q)^{-1} Q^t V = \lambda \qquad\qquad (2.2.6)$$

Le vecteur résidu de projection dans la base tronquée à m<n vecteurs est déterminé par :

$$R(m) = V - \lambda \sum_{i=1}^{m} \lambda_i \, Q_i \qquad\qquad (2.2.7)$$

L'expression du scalaire résidu de projection est donnée par

$$r(m) = \|R(m)\| \qquad\qquad (2.2.8)$$

**Propriété métrique des modes**
L'utilisation de la norme infinie (telle que le plus grand déplacement de $Q_i$ est égal à 1) permet d'obtenir une valeur métrique des coefficients modaux. C'est-à-dire que chaque coefficient modal $\lambda_i$ est attribué d'une valeur exprimée en mètres qui donne la contribution du mode $Q_i$ à la forme V (vecteur défaut en principe).

**Troncature**
La projection dans la base complète n'a aucun intérêt car elle donnera autant de paramètres de formes modales que de données de mesure. On projette donc dans une base tronquée aux modes de plus basses fréquences (donc de plus faibles complexités). On peut alors observer l'évolution du résidu avec le nombre de modes pris en charge.

Dans [FOR 07], nous avons présenté l'application de la méthode de décomposition modale à l'analyse d'un défaut mesuré sur un piston de moteur thermique. La figure 2.2.1 montre le processus de traitement de la mesure (a) à l'obtention des coefficients modaux métriques $\lambda_i$ (c). La dernière image représente la reconstruction de la forme à l'aide des 7 modes les plus contributifs (ceux qui font diminuer le résidu de façon la plus importante). Nous cherchons une description de la géométrie actuelle aussi fine que possible (résidu minimal) mais avec le plus petit nombre de paramètres ($\lambda_i$).

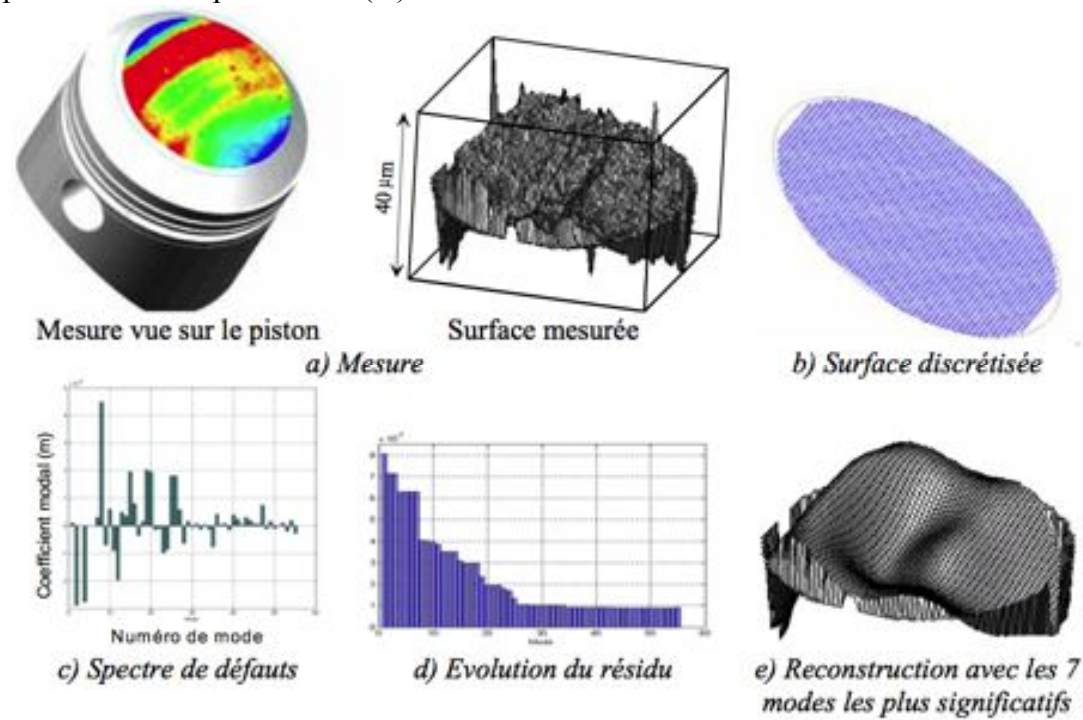

*Figure 2.2.1 De la mesure à la projection modale d'un défaut*

**Remarques**
- La base modale naturelle n'est pas orthogonale au sens euclidien mais au sens matriciel.
- La méthode présentée ici est la dernière évolution de nos travaux. La première modification a été d'introduire les coefficients métriques puis nous avons utilisé une orthonormalisation des vecteurs propres. Ces évolutions ont été liées à des applications industrielles.

## 2.2.1 Possibilités d'un multi matériau

La mise en œuvre d'une modélisation experte ouvre de nombreux champs, à l'instar de la démarche d'analyse des structures. L'utilisation de plusieurs matériaux permet de concevoir un modèle générant des formes propres adaptées à un besoin particulier (amplifications des amplitudes de défauts non équiréparties sur une géométrie,…).
Dans bien des cas, il peut être intéressant de définir un comportement qui rompt la symétrie comportementale. Il est aisé de définir un tel comportement. La figure 2.2.2.a montre un exemple de géométrie qui sert de base à cette démarche. Il nous est possible de définir un maillage de surface à base de poutres (des coques à matériau orthotropes auraient le même comportement) dont les caractéristiques sont dissymétriques (figure 2.2.2.b). Une autre possibilité est d'exploiter une anisotropie matériau en utilisant dans l'outil de modélisation des structures en matériaux composites.

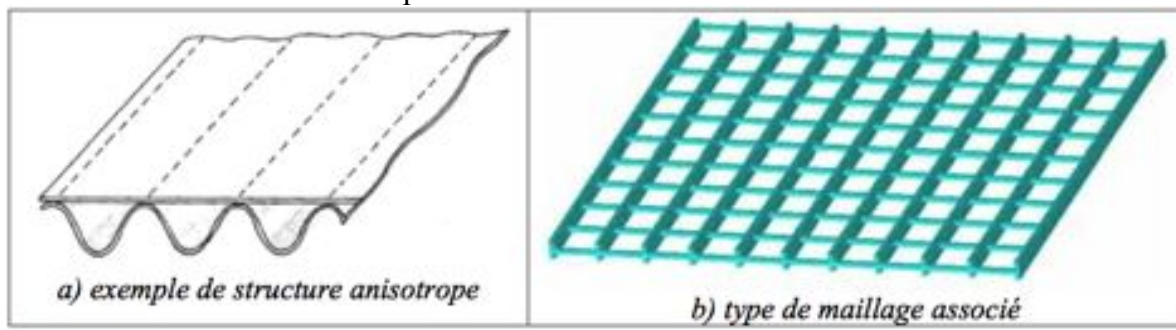

*a) exemple de structure anisotrope*          *b) type de maillage associé*

*Figure 2.2.2   Modèle d'anisotropie structurale*

*Les premiers modes naturels*
(

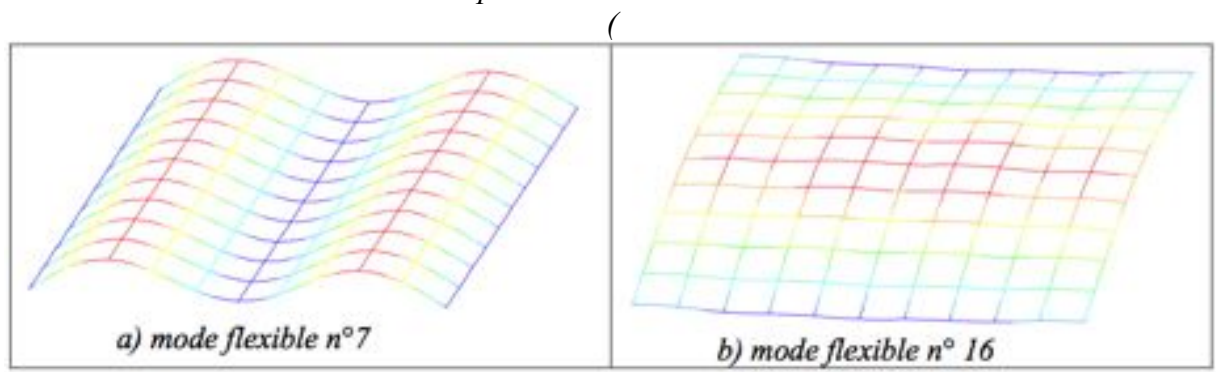

*a) mode flexible n°7*          *b) mode flexible n° 16*

figure 2.2.3) privilégient naturellement les plus grandes souplesses. Le septième mode flexible a quasiment deux périodicités dans le sens « souple » alors qu'il faut attendre le seizième pour avoir la première forme dans le sens « raide » du modèle.

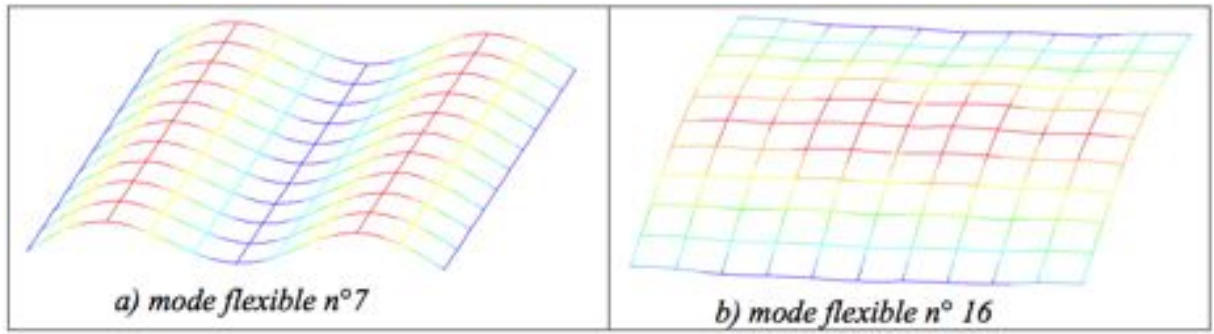

*Figure 2.2.3  Modes d'une plaque rectangulaire anisotrope*

Une utilisation de ce type de modélisation peut être l'étude des ondulations d'usinage où une direction donnera de plus grandes longueurs d'onde qu'une autre. Une coopération avec l'Université Technologique de Kaunas en Lituanie est en cours sur ce thème. Ce projet de deux ans (2007-2008) et liant six personnes sur des séjours de dix à vingt jours a pour objet l'étude des ondulations et rugosités 3D en relation avec les paramètres de procédé d'usinage.

## 2.3  Modélisations possibles

Nous avons fait le choix de ne présenter que deux types de comportements qui répondent de manière privilégiée aux études de variations de formes. La méthode des éléments finis ouvre une grande palette d'objets qui permettraient nombre de variations, néanmoins, il est intéressant de montrer comment avec peu d'outils on replie tous les besoins en analyse des défauts géométriques des courbes et des surfaces. Il est naturel d'utiliser les comportements de poutre pour décrire les courbes et les comportements coques pour décrire les surfaces. La richesse des comportements poutres et coques synthétise les comportements membrane et flexion, pour les courbes et les surfaces.

Les géométries nominales sont, soit de type bidimensionnelle, soit tridimensionnelle, le symbole « **X** » désigne les usages préférentiels du comportement considéré et le symbole « **\*** » les usages possibles complémentaires.

Il est évident que tout type d'élément peut convenir à la modélisation mais il est intéressant de voir qu'avec ces deux types d'éléments, nous pouvons balayer facilement l'ensemble des cas possibles. En incluant (ou pas) leur comportement à des mouvements de flexion, on peut prendre en compte (ou pas) les variations de dimensions en plus de celles de formes.

| Dimension | Type de géométrie | | Poutres | Coques |
|---|---|---|---|---|
| 2D | 1D | Segment | X | * |
| | | Cercle | X | * |
| | | Courbe | X | * |
| 3D | 1D | Segment | X | * |
| | | Cercle | X | * |
| | | Courbe | X | * |
| | 2D | Disque-ellipse | * | X |
| | | Rectangle | * | X |
| | | Quelconque | * | X |
| | 3D | Cylindre | * | X |
| | | Cône | * | X |
| | | Sphère | * | X |
| | | Quelconque | * | X |

*Tableau 2.3.1   Correspondances entre comportements et géométries*

**Remarques**
-     Le choix de discrétisation d'une surface en poutres peut être très judicieux dès lors qu'on désire décrire des défauts non isotropes. Les défauts étant orientés et de longueurs d'ondes différentes selon deux directions principales (fussent-elles locales) on peut alors utiliser des poutres de sections différentes dans les deux directions et même orienter le maillage selon les directions principales d'orthotropie de ces défauts.
-     Le spécialiste a à sa disposition un outil d'une grande richesse de modélisation qui lui permet de simplifier au maximum l'écriture des variations de géométrie selon le nombre de critères.

Dans la figure 2.3.1, les défauts de forme d'un accostage ou de la surface d'une pièce plastique peuvent être analysés selon deux types de modélisations. Dans (b) le modèle poutre permet de décrire l'accostage et dans (c) le modèle coque permet d'analyser les défauts de surface.

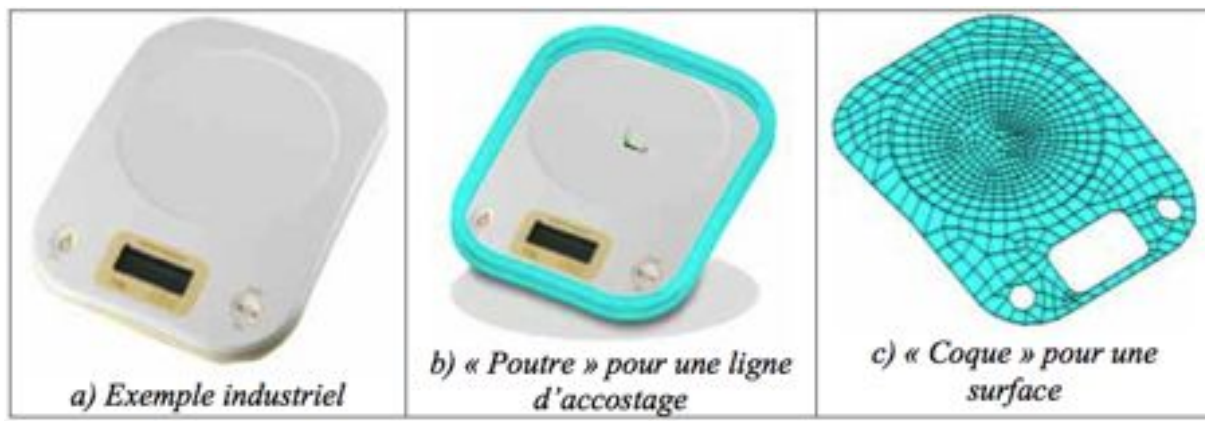

a) Exemple industriel     b) « Poutre » pour une ligne d'accostage     c) « Coque » pour une surface

*Figure 2.3.1 Modèle « poutre » ou modèle « coque »*

## 2.4  Surfaces

Nous pouvons décomposer les surfaces en deux catégories, les surfaces élémentaires et les surfaces gauches. Toute surface peut être décrite à l'aide d'une équation à deux variables mais cette mise en équations peut être paramétrable simplement (surfaces élémentaires) ou nécessiter des outils numériques. Nous nous intéressons aux variations de formes de ces surfaces, nous présenterons à ces fins des méthodes issues de la littérature.

## 2.4.1 Cas du disque

Nous avons comparé dans [FOR 07] le paramétrage modal aux polynômes de Zernike qui permettent de définir tous (paramétrage exhaustif) les défauts que peuvent avoir les disques, dans l'objectif de décrire une forme avec le minimum de grandeurs. Ces polynômes sont principalement appliqués à la décomposition des défauts en optique et font l'objet d'une norme spécifique ([ISO 10110-5]). Il est à noter que les polynômes (écrits en coordonnées polaires) de Zernike (2.4.1) sont proches des modes naturels d'un disque aux conditions aux limites libres. La fonction de déplacements $Z(\rho,\theta)$ représente l'écart d'un point par rapport à la référence.

$$Z_n^m(\rho,\theta)=\begin{cases} N_n^m R_n^{|m|}(\rho)\cos m\theta\,; pour\,m\geq 0 \\ -N_n^m R_n^{|m|}(\rho)\sin m\theta\,; pour\,m<0 \end{cases} \quad \text{où } \delta_{m0} \text{ est la fonction delta de kroneker et} \quad (2.4.1)$$

$$N_n^m=\sqrt{\frac{2(n+1)}{1+\delta_{m0}}} \qquad\qquad R_n^{|m|}(\rho)=\sum_{s=0}^{(n-|m|)/2}\frac{(-1)^s(n-s)!}{s!\,[0,5(n+|m|-s]!\,[0,5(n-|m|-s]!}\rho^{n-2s}$$

Les modes naturels d'un disque sont les solutions analytiques de (2.1.1), (2.1.2) et (2.1.3). Dans le cas des conditions aux limites du type « bords libres », les solutions analytiques, données par [LEI 69] existent mais sont complexes à exploiter aussi, il nous est plus aisé d'utiliser celles données par la MEF (modes de la figure 2.1.1).

Si nous reprenons le cas de la surface supérieure du piston présenté page 11 et que nous comparons la projection modale à celle des polynômes de Zernike, nous obtenons les deux spectres de la figure 2.4.1. Dans le cas de l'exemple traité, nous constatons que le paramétrage de Zernike demande plus de coefficients que le modal.

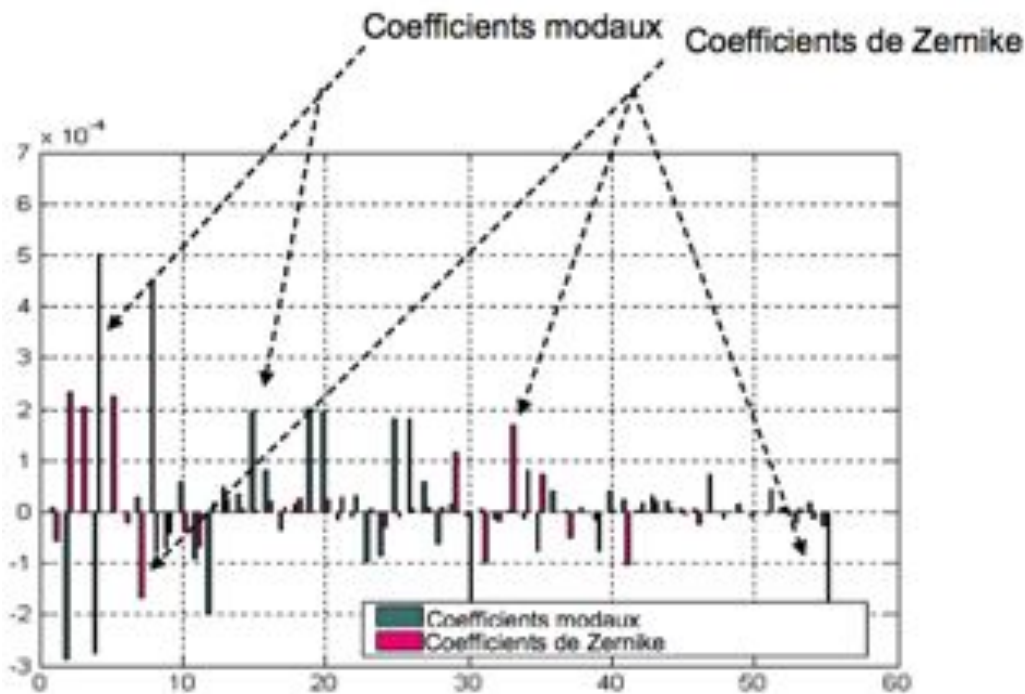

*Figure 2.4.1 Comparaison Zernike/Modes naturels*

## 2.4.2 Rectangle en modèle membrane

Dans le cas d'une surface rectangulaire, modélisée à l'aide d'une membrane tendue à bords libres, nous obtenons les solutions en déplacements suivantes :

$$Z_{pq}(x,y) = \cos\left(\frac{p\,\pi\,x}{Lx}\right)\cos\left(\frac{q\,\pi\,y}{Ly}\right) \tag{2.4.2}$$

Lx et Ly sont les longueurs des côtés du rectangle dans les axes x et y. Les entiers p et q sont des paramètres d'ordre de formes. Nous pouvons noter que cette équation est celle qui régit la base Discrete Cosine Transform (DCT). Elle a pour avantage d'être très polyvalente comme le montre ses applications en traitement du signal (compressions d'images en particulier). Huang l'a employée pour identifier les défauts de forme dans une pièce emboutie [HUA 02].

**Remarques**
De la même manière, les cordes tendues donnent des solutions identiques aux séries de Fourier. L'utilisation des modèles sans raideur intrinsèque de flexion (cordes, membranes) donne généralement des modes propres de formes mathématiques plus simples (formes harmoniques).

# 3 Défauts technologiques

Dans le cadre de contrats de recherche, le besoin de nos partenaires industriels était d'analyser des défauts dont certains étaient clairement définis. La base modale est certes exhaustive dans sa capacité à décrire les formes mais les modes naturels ne sont pas forcément des formes qu'on recherche. Nous avons donc introduit des modes technologiques. Il est alors nécessaire de recalculer une base modale prenant en compte ces modes et le résidu de projection.

**Définitions :**
- Les modes naturels sont les modes calculés initialement.
- Les modes technologiques sont les formes introduites spécifiquement dans la base modale.
- La base reconstruite est définie avec les modes naturels et les modes naturels recalculés.
- Une base technologique peut être définie seulement à l'aide de défauts technologiques.

## 3.1 Cas d'un cylindre-tiroir

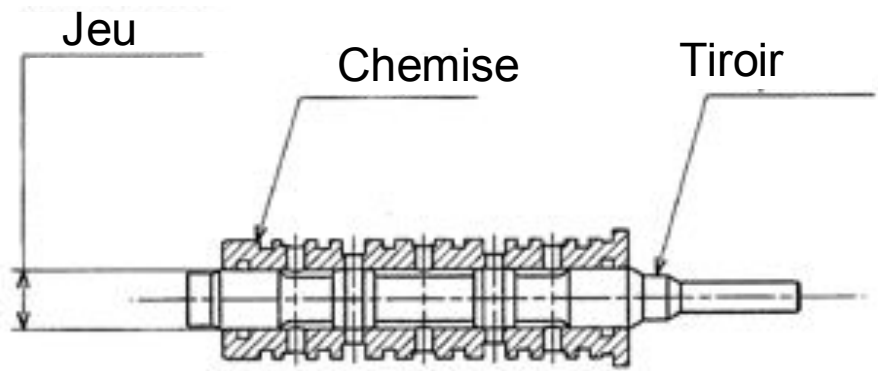

*Figure 3.1.1 Distributeur hydraulique*

Nous nous sommes intéressés à la forme du tiroir de ce distributeur. Les défauts à trier étaient, selon l'industriel, la taille (appairage), la conicité et le défaut « en banane ». Nous avons effectué [ADR 07a] des mesures puis des analyses de ce tiroir. Les défauts recherchés sont dans la base modale naturelle mais sous la forme d'une combinaison de modes naturels. Il serait possible de redéfinir un modèle permettant d'obtenir ces défauts à l'aide d'une démarche d'optimisation. Le plus cohérent est de les intégrer dans la base modale dans la mesure où ils sont indépendants et de recalculer les modes naturels en conséquence pour garder la structure de la base modale.

La fonction à analyser est l'assemblage du tiroir dans le corps du distributeur, les quatre cylindres doivent être vus comme étant des portions d'un seul cylindre. Le modèle d'analyse (Figure 3.1.2) est alors un cylindre complet (b). Les mesures ne pouvant être faites que sur les portions, la projection (cf. équation (2.2.6) page 10) du défaut dans la base modale est faite sur le sous modèle (c) réduit à quatre cylindres. Les conditions aux limites sont du type libre pour permettre d'extraire la position des mesures.

**Remarques**
- Un modèle poutre cylindrique sur l'axe permet d'extraire les défauts de rectitudes.
- Un modèle de quatre cylindres indépendants permet d'analyser les défauts de position des cylindres les uns par rapport aux autres et leurs défauts propres.

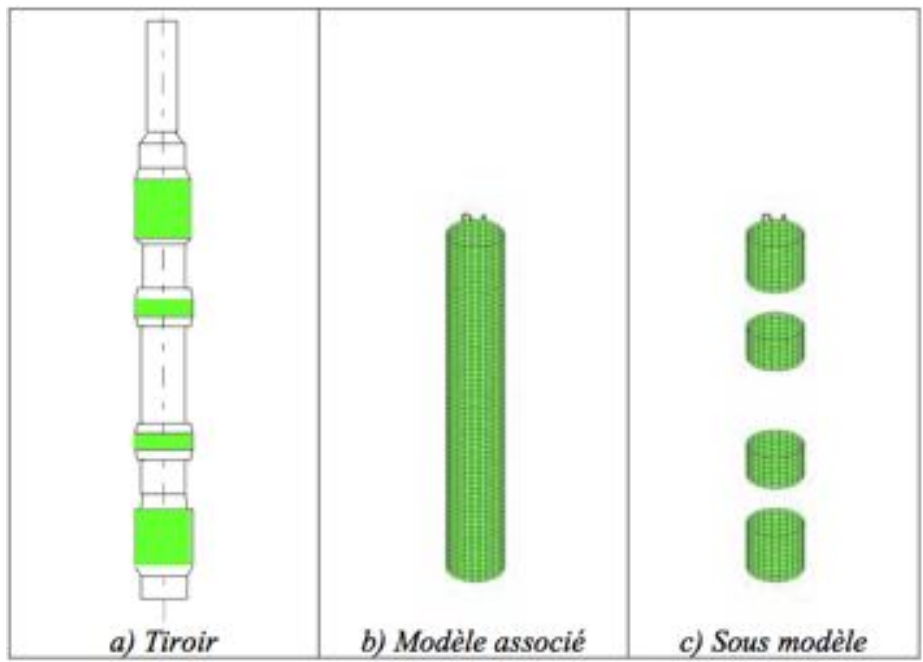

*Figure 3.1.2 Modèle d'analyse du tiroir de distributeur hydraulique*

La figure 3.1.3 montre des modes naturels de la structure associée aux quatre portions de cylindres du tiroir de distributeur. Les couleurs, du bleu vers le rouge sont purement indicatives. Ces formes sont présentées de façon groupée :
- Les deux premiers (a) sont des modes rigides de décalage radial et de rotation.
- Les quatre suivants (b) sont des modes de « flexion » qui permettent d'analyser les défauts de cylindricité globale à circularité parfaite.
- Les modes en forme de « tonneau » (c) sont des modes de cylindricité d'ordre local.
- Les formes d'ovalité de différents ordres sont présentées en (d).
-

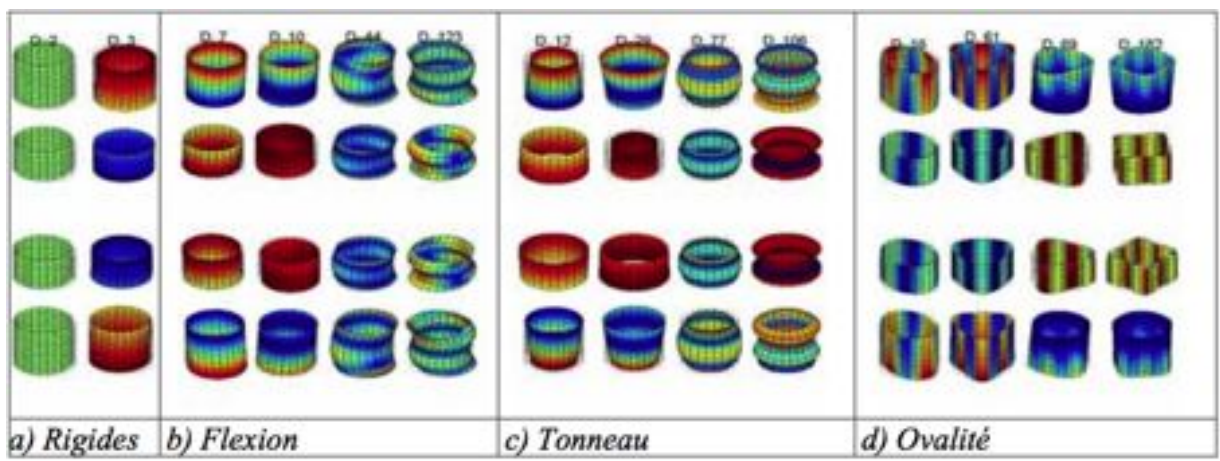

*Figure 3.1.3 Modes naturels d'un tiroir de distributeur (décomposé en quatre cylindres liés)*

**Remarques**
- Les modes rigides ne devraient pas être nécessaires car l'appareil de mesure (mesureur de forme Mahr MMQ40) devrait les rejeter, néanmoins nous avons observé des écarts résiduels de position.

- Dans la base des modes naturels d'un cylindre à parois minces, il n'y a pas de forme de dilatation radiale simple qui permette d'identifier le défaut de taille, ni de dilatation proportionnelle à la position axiale qui permette d'identifier la conicité. Ces formes sont des combinaisons linéaires des modes naturels. Pour observer spécifiquement ces formes, il faut les introduire comme « modes technologiques » (Figure 3.1.4).

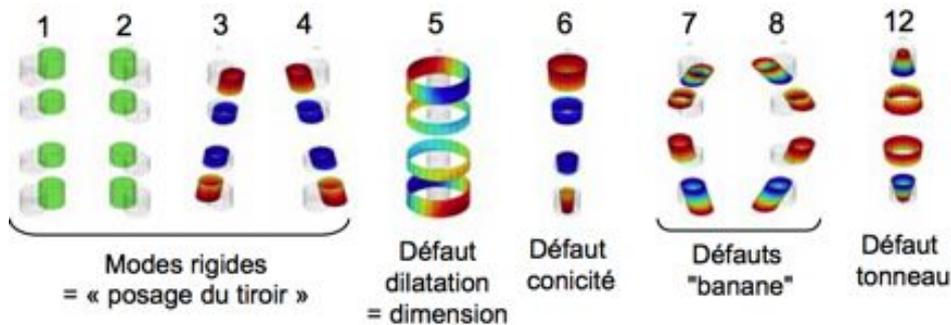

*Figure 3.1.4 Modes technologiques du tiroir de distributeur*

L'introduction de ces modes technologiques demande à modifier la base modale en projetant les modes naturels. Le calcul des coefficients modaux d'un tiroir dans la base reconstruite se fait de la même manière que pour la base naturelle. Le résultat peut être mis sous la forme d'un spectre (Figure 3.1.5). Le plus grand défaut (n°5) serait la taille mais elle ne pose pas de problème, les diamètres étant appairés. La position est donnée pas les quatre premiers modes et la conicité (n°6) dépasse 1µm. Les autres défauts sont faibles.

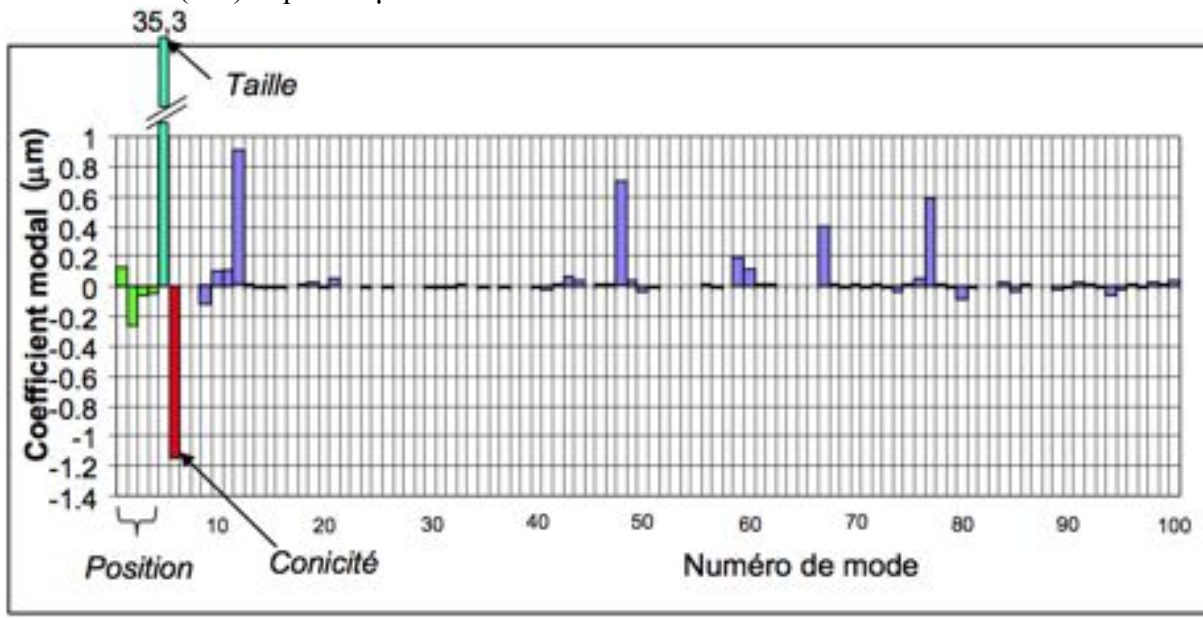

*Figure 3.1.5 Coefficients modaux du tiroir de distributeur*

L'utilisation du logiciel de métrologie d'Eric Pairel [PAI 05] [PAI 07] associant un calibre virtuel (cylindre ou cône) à l'ensemble des points nous a permis de valider nos mesures de taille et de conicité.

La représentation spectrale est intéressante car elle met en évidence les ordres de grandeur des différents défauts de l'objet mesuré. Dans l'analyse d'une série de pièces il est pertinent de reconnaitre l'écart aléatoire du systématique. L'étude d'un lot de pièces est présentée ci-après.

# 4  Etude d'un lot de formes

Dans le cadre du contrat interregIII « tolérancement des systèmes assemblés », les aspects statistiques sont très importants (tolérancement inertiel) [ADR 06b] [PIL 08] [ADR 07a]. Nous avons développé les analyses de formes statistiques dans ce contexte en l'appliquant à des cas industriels [ADR 06a]. Nous présentons ci après deux applications industrielles de nos travaux.

## 4.1  Lot de tiroirs de distributeurs

**Problématique**
Les spécifications du bureau d'études (client interne) sont écrites selon les normes mais la volonté du site de production est de réaliser des formes les plus fonctionnelles possibles. L'entreprise partenaire voulait connaître les formes et leurs dispersions de l'ensemble du processus de production des tiroirs de distributeurs. L'étude d'un lot (réduit mais caractéristique) nous permet de montrer les formes technologiques mais aussi le résidu de ces formes.

**Démarche**
Nous avons analysé un lot de dix tiroirs distributeurs. Nous avons mis en évidence les défauts de taille et de conicité (Figure 3.1.5) qui sont prépondérants. Dans la figure 4.1.1 nous montrons le résidu (les modes 5 et 6 ont été retirés) de ces défauts pour un lot de tiroirs. Il apparaît tout d'abord que le procédé est relativement stable, en effet, sur 200 modes (dont 100 sont représentés ci dessous) il reste 4 modes prépondérants (12, 45, 67 et 77). L'entreprise partenaire peut effectuer son expertise sur la criticité de ces formes.

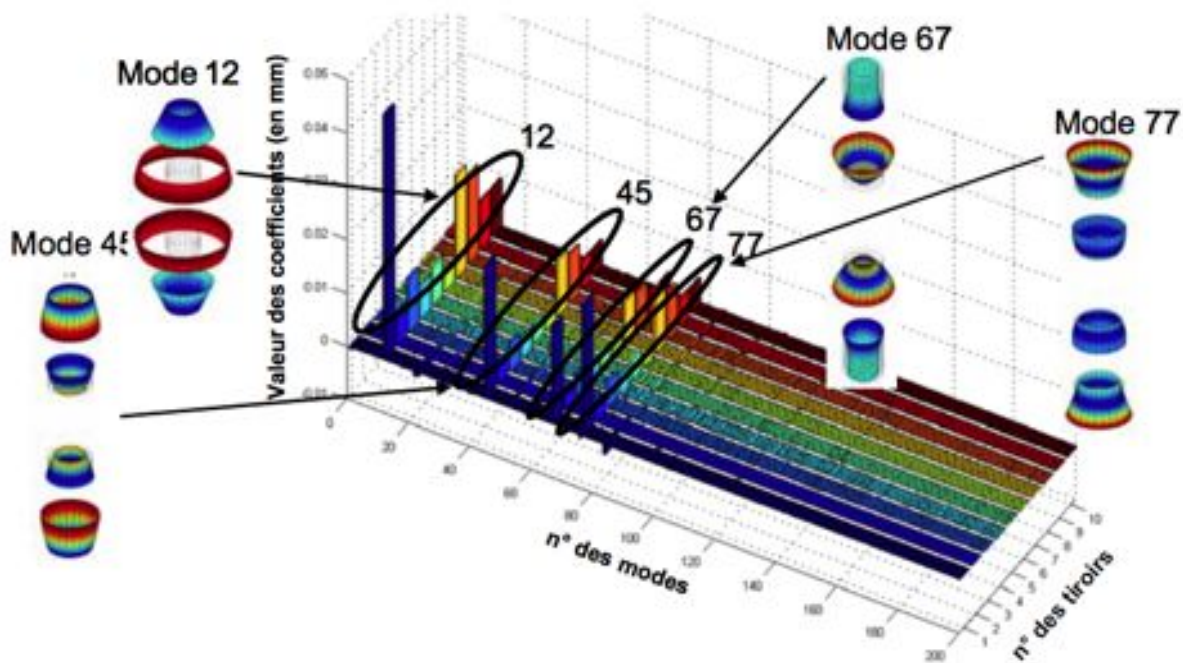

*Figure 4.1.1 Signature modale résiduelle d'un lot de tiroirs de distributeur*

**Remarques**

- Les quatre formes (12, 45, 67 et 77) peuvent être décrites par une forme technologique si l'expert la relie à une fonction.
- Il est possible d'aller plus loin et de montrer (comme on le fait dans le paragraphe suivant) la forme moyenne lissée du lot ou le résidu lissé.

## 4.2 Lot de capots de balance

**Problématique**

L'entreprise TEFAL-ARTICLES DE PESAGE met en œuvre un plan de surveillance des pièces plastiques à l'aide d'un tableau de caractéristiques. Ce tableau n'est pas simple à lire (beaucoup de données) et ne permet pas de voir les couplages de données.

**Démarche**

En prenant un lot de capots (5 seulement) de balance, nous décrivons sa signature modale et en extrayons les caractéristiques moyennes et écarts types. Cette démarche permet aux industriels d'analyser avec une méthode rigoureuse et visuelle les défauts (modes naturels) de forme d'un lot de pièces. Notre collaboration a permis de proposer [FAV 07a] [PIL 06] plusieurs modes de représentations d'un lot dont la figure 4.2.1 montre un exemple.

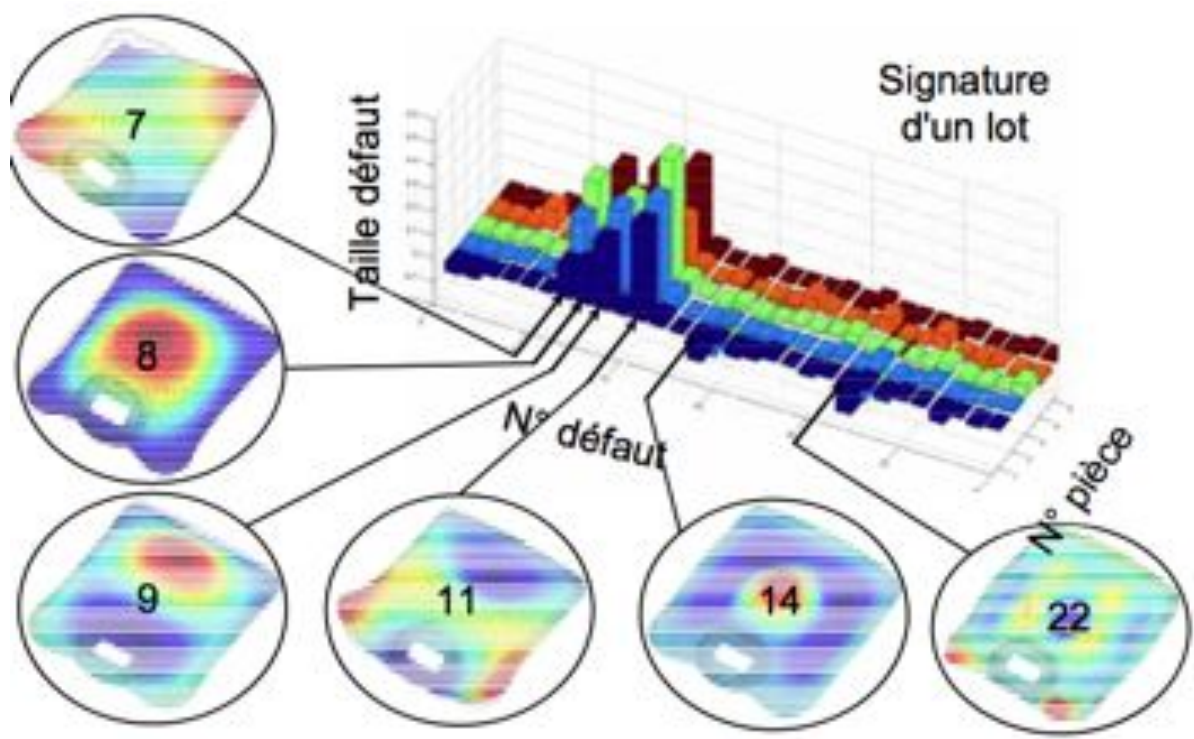

*Figure 4.2.1 Signature modale d'un lot de capots de balance*

Ce lot de balances peut être représenté sous la forme d'un double spectre (figure 4.2.2) de valeurs moyennes et d'écarts types.

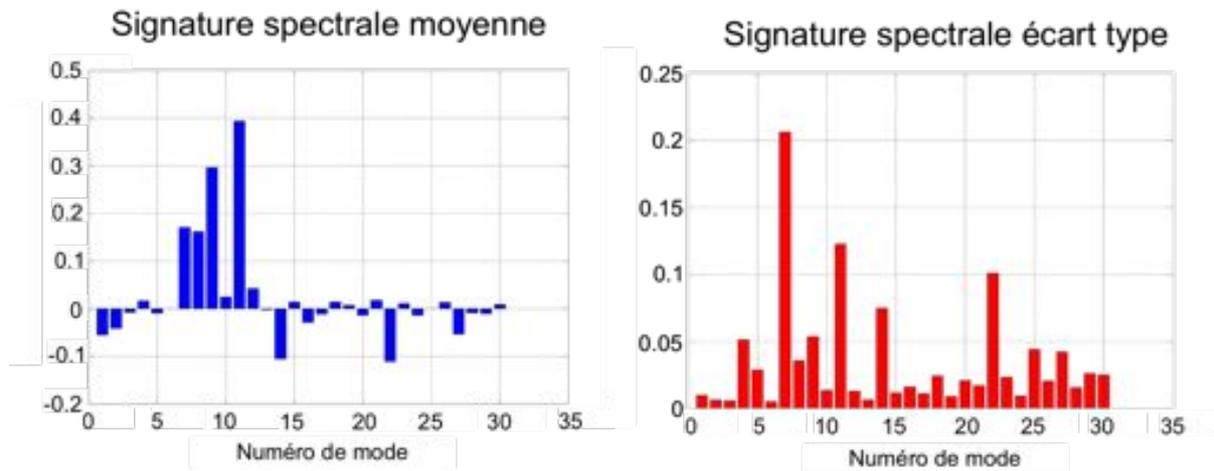

*Figure 4.2.2 Signature statistique d'un lot de capots de balance*

On effectue alors une troncature du spectre qui permet d'obtenir la forme moyenne du lot de balances et la cartographie des écarts types. La figure 4.2.3 montre la correspondance entre les deux représentations et une gaussienne en un point particulier. On peut voir le défaut moyen et son amplitude ainsi que les zones dans lesquelles on aura les plus grandes dispersions.

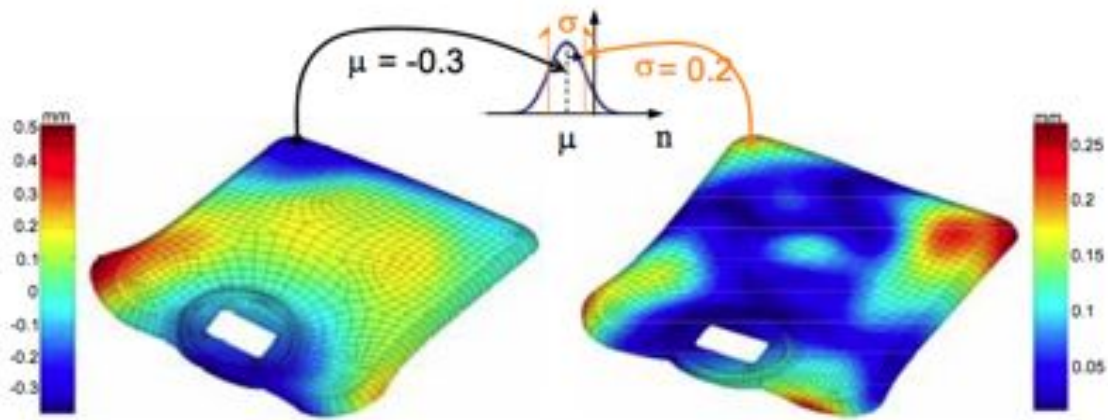

*Figure 4.2.3 Moyenne 3D et écart type 3D d'un lot de balances*

# 5 Etude d'assemblages avec défauts de forme

## 5.1 Assemblage 2D

### 5.1.1 Introduction

Les travaux les plus récents [ADR 07] [ADR 08] portent sur la problématique de l'assemblage des surfaces avec défauts de forme. L'assemblage se fait avec deux notions importantes, la définition d'une action $T_{1A2}$ (torseur d'efforts) de liaison et la construction d'une enveloppe convexe de contact. Nous proposons un assemblage simple (Figure 5.1.5) dans lequel nous considérons les surfaces Bi parfaites (pour raccourcir l'étude). Cf est la condition fonctionnelle à respecter entre les surfaces Bi.

## 5.1.2 Simple assemblage 2D

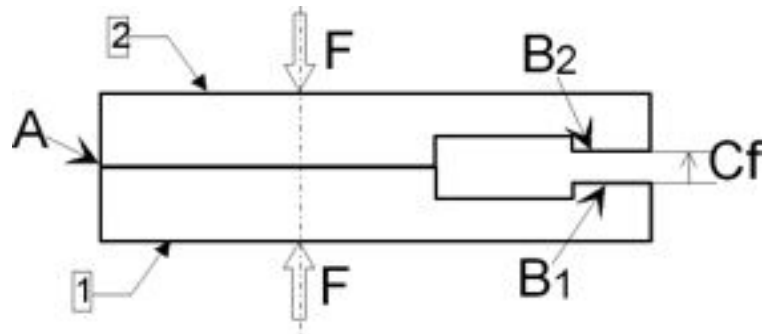

*Figure 5.1.1 Assemblage 2D*

Le modèle géométrique permettant de faire le paramétrage modal est une poutre dont la longueur est la longueur de contact. La figure 2 montre les six premiers modes.

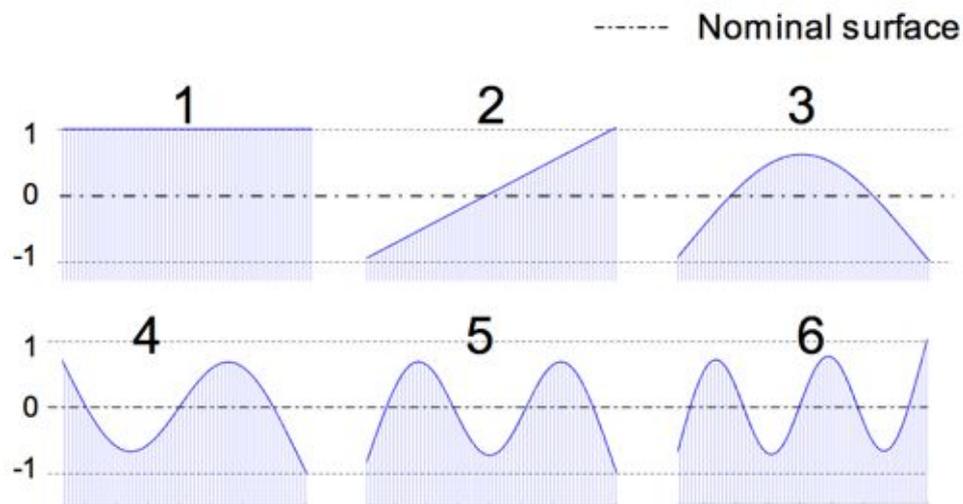

Fig. 2 Modes propres de la ligne de contact

La figure ci-dessous montre comment une forme peut être paramétrée à l'aide des modes propres. On obient alors un filtrage géométrique et les niveaux des modes les plus influents.

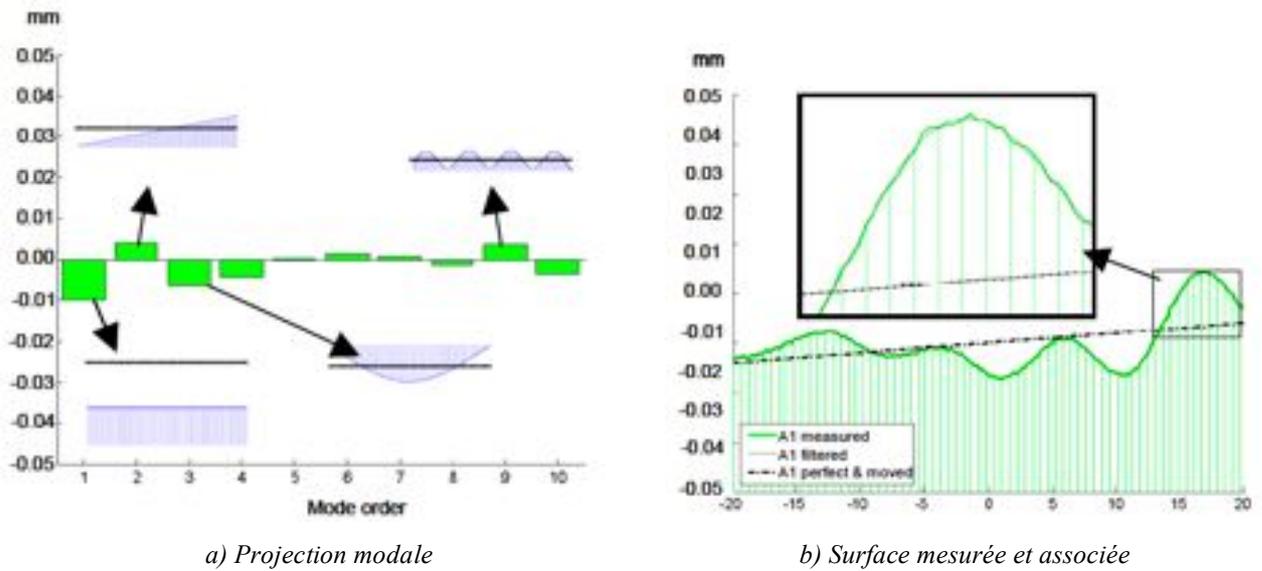

*a) Projection modale*                    *b) Surface mesurée et associée*

Fig. 3 Paramétrage modal

La relation suivante donne les coeffients modaux de la surface différence.

$$\lambda_{1A2} = \lambda_{A2} - \lambda_{A1} \hspace{5cm} (3)$$

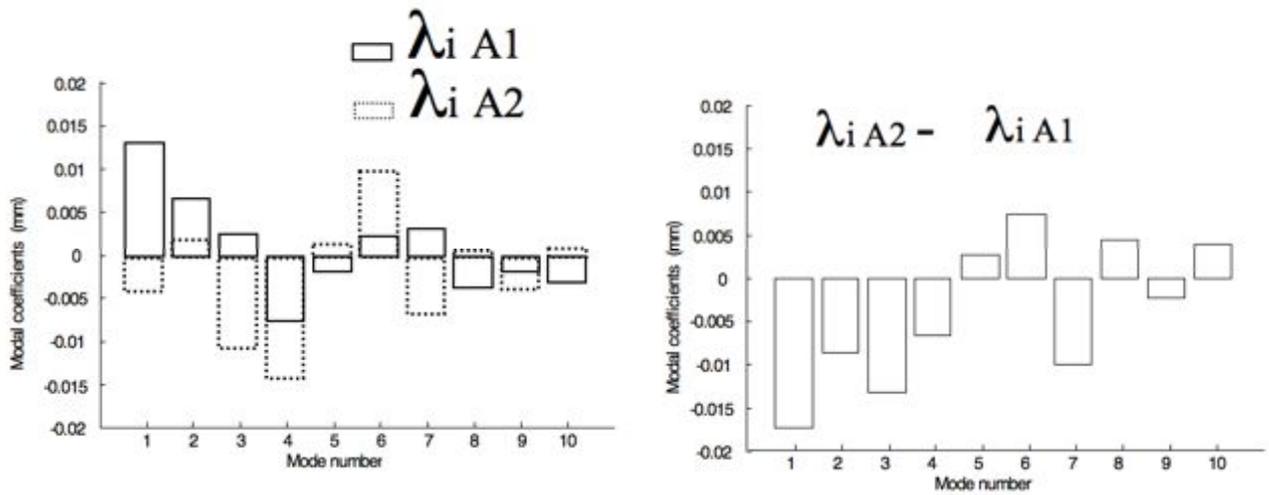

Fig. 4 Coefficients modaux de la surface différence

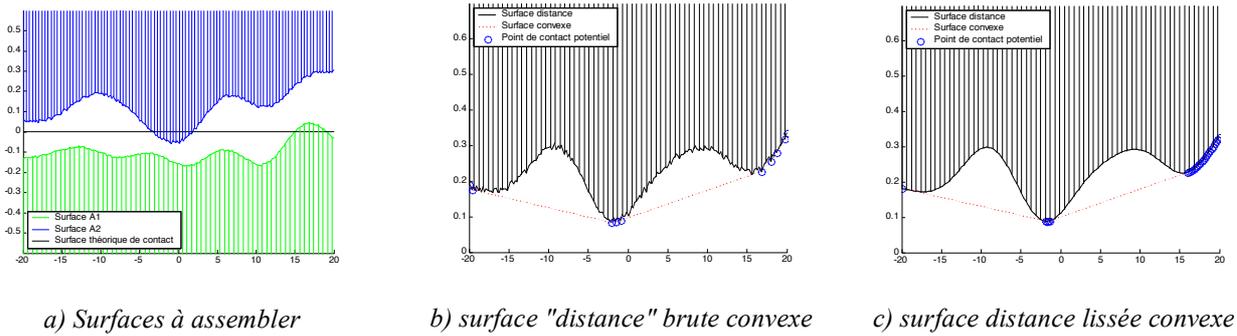

*a) Surfaces à assembler*     *b) surface "distance" brute convexe*     *c) surface distance lissée convexe*

*Figure 5.1.5 Détermination de la surface convexe « distance de contact »*

La figure 5.1.5 montre comment, à partir de la définition des surfaces à assembler nous déduisons la surface distance puis la surface convexe associée. Le paramétrage modal sert à effectuer un lissage et à déterminer la surface distance convexe. En appliquant à cette dernière et à son vis-à-vis un torseur de liaison (eq. 4) , on trouve alors la configuration de contact. Elle est exprimée en paramètres modaux puis traduite en composantes de TPD de manière à être utilisée dans une étude plus générale de l'assemblage.

$$
\begin{Bmatrix} \lambda_{R1} \\ \lambda_{R2} \\ \lambda_{R3} \\ \lambda_{R4} \\ \lambda_{R5} \\ \lambda_{R6} \end{Bmatrix} = \begin{bmatrix} \alpha_{11} & \alpha_{12} & \ldots & \alpha_{16} \\ \alpha_{21} & \alpha_{22} & \ldots & \alpha_{26} \\ . & . & \ldots & . \\ . & . & \ldots & . \\ \alpha_{61} & \alpha_{62} & \ldots & \alpha_{66} \end{bmatrix} \begin{Bmatrix} Tx_A \\ Ty_A \\ Tz_A \\ Rx_A \\ Ry_A \\ Rz_A \end{Bmatrix} \qquad (4)
$$

Comme les 6 premiers modes sont des modes rigides (3 dans un mouvement plan) , nous écrivons un opérateur linéaire *[$\alpha_{ij}$]* dans le but de repréenter les relations entre les coefficients modaux $\lambda_{Ri}$ dans les composantes du TPD correspondant (4).

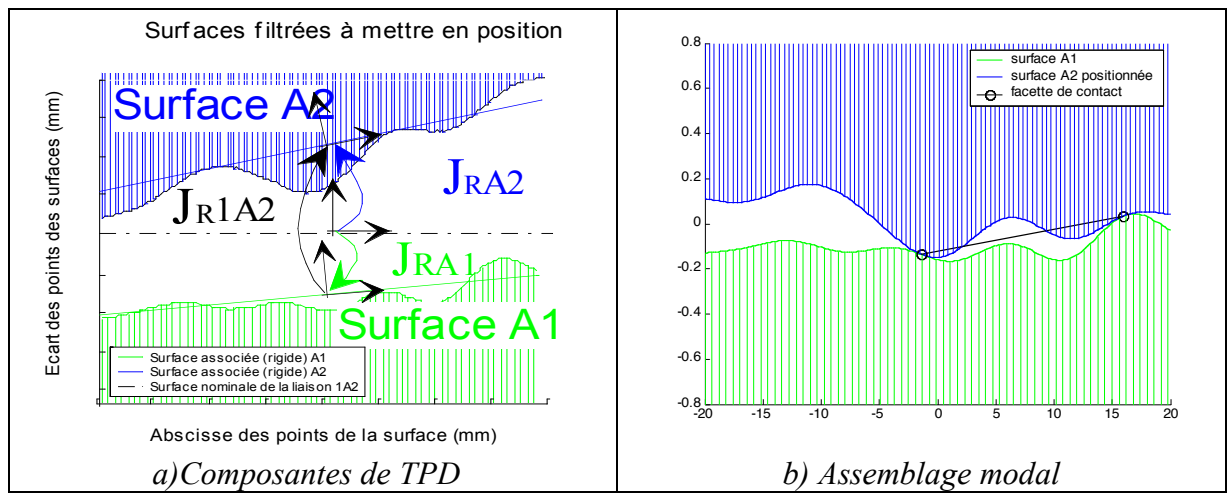

*a)Composantes de TPD*          *b) Assemblage modal*

*Figure 5.1.6 Détermination de la surface distance de contact convexe*

La figure 5.1.6 montre comment on peut définir $J_{RiXj}$ comme étant le torseur jeu rigide (sans défaut de forme) entre les surfaces associées sans défaut de forme et $J_{FiXj}$ celui qui est spécifique à la forme. On obtient alors par addition, $J_{iXj}$, le torseur de la liaison X entre les pièces i et j.

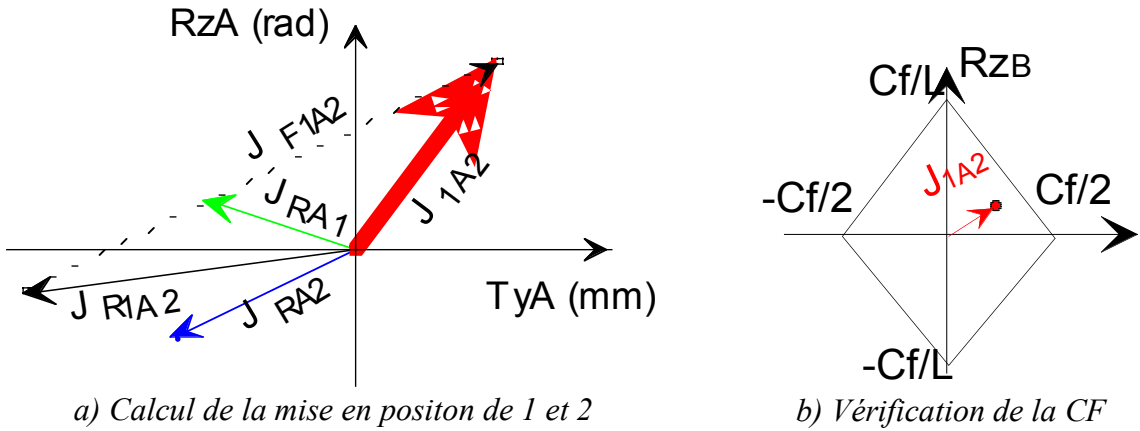

*a) Calcul de la mise en positon de 1 et 2*          *b) Vérification de la CF*

*Figure 5.1.7 TPD résultant et CF*

Dans la figure 5.1.7, la somme du torseur jeu des surfaces associées sans défauts de formes ($J_{R1A2}$) avec le celui dû aux défauts de formes assemblés ($J_{F1A2}$) donne la mise en position $J_{1A2}$ sous l'action de contact $T_{1A2}$. En transportant ce torseur en B, on vérifie que ses composantes sont circonscrites au domaine spécifié.

### 5.1.3  Lot d'assemblages 2D

Dans [I_ADR 08], nous abordons les aspects statistiques d'assemblage avec défauts de forme. Afin de prendre en compte les écarts moyens et les dispersions dans les distributions de défauts nous utilisons le concept « d'inertie » développé par Maurice Pillet en l'élargissant à des caractérisations de formes [ADR07].

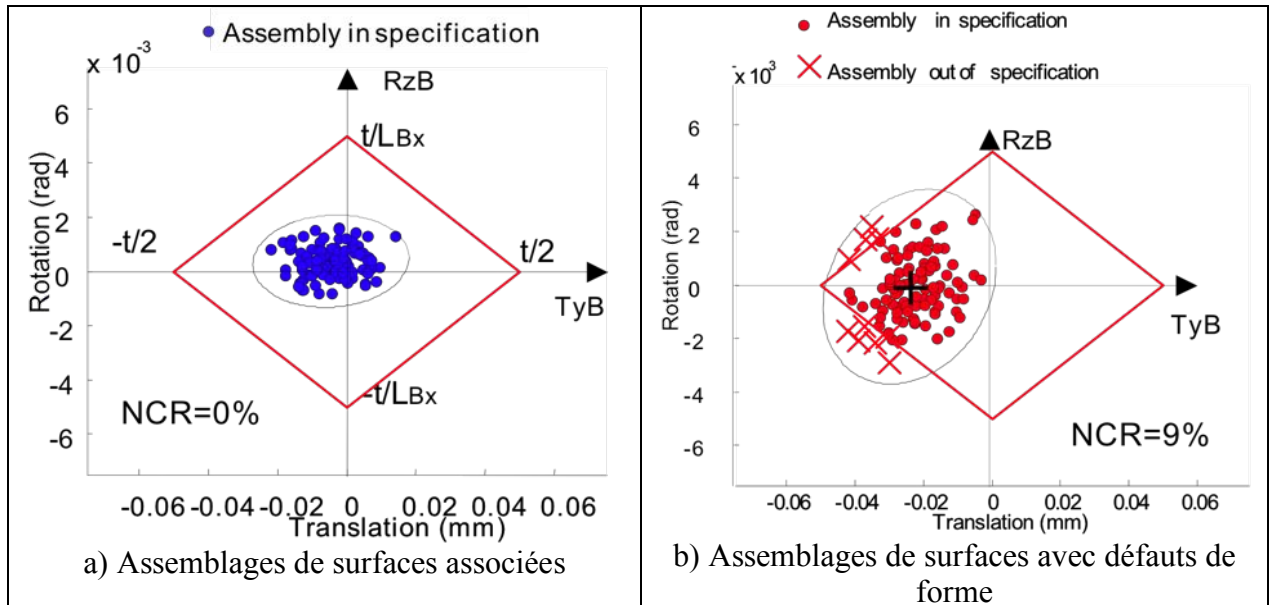

a) Assemblages de surfaces associées          b) Assemblages de surfaces avec défauts de forme

**Fig.  8 Étude de la conformité d'un lot de 100 assemblages**

On montre par des simulations, dont la prise en compte du défaut de forme peut changer considérablement le taux de non-conformité. Nous montrons surtout comment le calculer à travers une méthodologie non-empirique.

### 5.1.4  Assemblage 3D

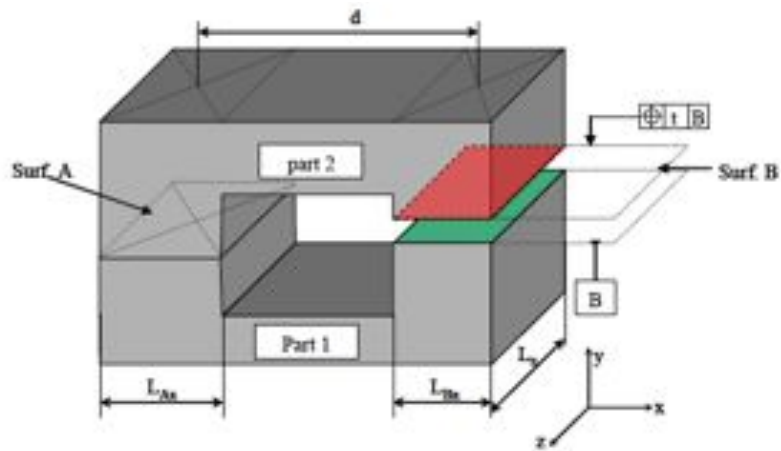

*Figure 9 Tolérancement de l'assemblage.*

#### 5.1.4.1   Simple assemblage

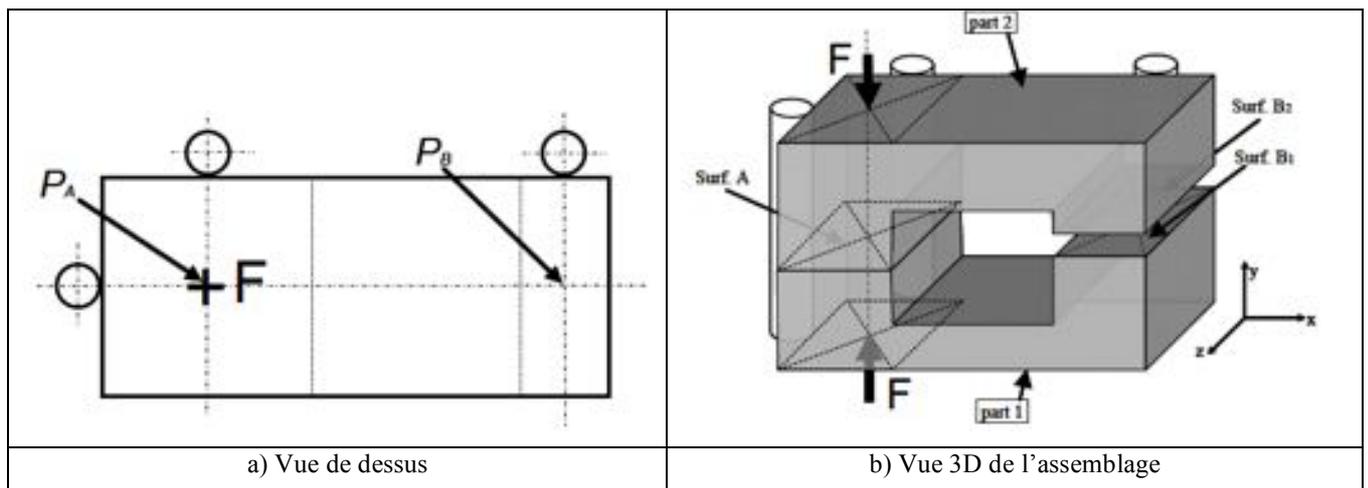

| a) Vue de dessus | b) Vue 3D de l'assemblage |

Figure 10 Construction de l'assemblage

#### 5.1.4.2   Lot d'assemblages

## 5.2  Qualité perçue des assemblages

Le besoin client n'est plus seulement simplement fonctionnel : aux exigences conventionnelles (assemblabilité, fonctionnalités basiques), s'ajoutent de nouvelles demandes en matière d'aspect. Il est nécessaire de définir une métrologie de l'aspect, on parle alors de qualité perçue. Celle-ci peut s'exprimer dans différents domaines, chacun associé aux sens de l'humain. Nous nous intéresserons aux seuls aspects géométriques.

Certains travaux utilisent la CAO comme solveur d'analyse d'accostage [WIK 03], mais la lenteur des résolutions limite le nombre de calculs. Nous avons choisi de développer une maquette logicielle dans une application de résolutions numériques. Dans [ADR 07], nous présentons le principe d'analyse d'un accostage avec défaut de forme en traitant en exemple, un cas industriel.

## 5.2.1  Les défauts d'accostage

Nous nous sommes déjà intéressés dans le chapitre 2 (accostage aile/projecteur) à cet aspect de la qualité géométrique en négligeant les défauts de formes, mais ici, nous complétons ces travaux en prenant en compte les défauts des bords des surfaces en contact. L'accostage décrit comment deux surfaces sont "cousues" l'une à l'autre. Il existe 6 défauts d'accostage que nous avons associés au Torseur de Petits Déplacements (TPD) d'une surface en prenant l'autre en référence. Nous sommes confrontés quotidiennement à une analyse d'accostages, de l'observation d'un carrelage, d'un assemblage de pièces plastiques, des éléments de carrosseries de voitures,…

### 5.2.1.1  Définition des 6 défauts d'accostage

Les accostages sont des défauts qui se décomposent, selon le TPD en trois translations et trois rotations, mais surtout du point de vue client-fournisseur en deux catégories à savoir les jeux et les affleurements. Les jeux sont les déplacements qui modifient la distance dans le plan local tangent aux surfaces, ils sont observés selon une direction normale à ce plan. Les affleurements leur sont complémentaires et sont observés selon les plans tangents.

Nous proposons la définition d'un repère local qui permette de décrire les accostages en discriminant les jeux et les affleurements le long d'une courbe de « couture »de deux surfaces. Une des deux surfaces doit être définie comme élément de référence. C'est à partir de cette dernière que les jeux et les affleurements sont définis.

En chaque point P de la courbe C d'intersection de deux surfaces $S_i$ et $S_j$, nous définissons le vecteur $\vec{x}$ tel que $(P, \vec{x})$ est tangent à la courbe C d'intersection des surfaces. Le vecteur $\vec{z}$ est tel que $(P, \vec{y})$ est dirigé selon la direction de spécification du jeu. Le vecteur $\vec{z}$ est tel que $(P, \vec{z})$ est dirigé selon la définition de l'affleurement. Il est de plus nécessaire de définir le point de vue de l'observateur qui permet de discriminer l'affleurement du désaffleurement selon qu'on voit ou pas l'interstice laissé par le décalage selon z de $S_2$ par rapport à $S_1$. Nous proposons comme convention que $\vec{y}$ soit dans le sens du jeu positif et $\vec{z}$ dans le sens affleurement.

La figure 5.2.1 met en évidence un repère d'accostage de deux surfaces $S_1$ et $S_2$. S1 est la surface de référence pour la définition des jeux et affleurements. Dès lors $(P, \vec{y})$ est tangent à $S_1$ et normal à $(P, \vec{x})$.

**Remarque :**
Si nous permutons les surfaces $S_1$ et $S_2$, les axes ne subissent pas de permutations mais peuvent tourner si les surfaces liées ne sont pas de tangence commune.

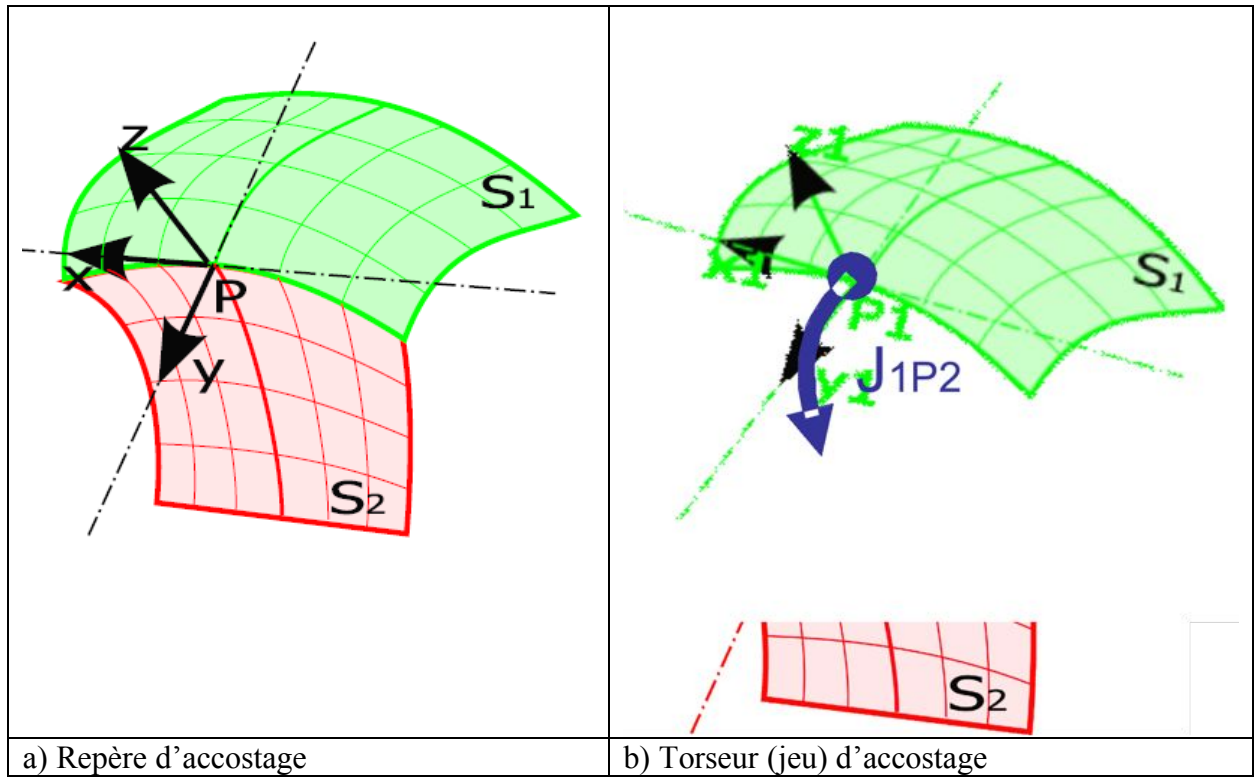

| a) Repère d'accostage | b) Torseur (jeu) d'accostage |

*Figure 5.2.1 Repère et torseur daccostage*

Chaque repère $(P, \vec{x}, \vec{y}, \vec{z})$ est orthonormé. L'ensemble de ces repères définit le domaine de définition des accostages des deux surfaces $S_i$ et $S_j$. Les points de mesure seront pris dans cet ensemble. Les deux surfaces ne sont pas nécessairement connectées, dans ce cas, les surfaces seront projetées afin de construire la courbe d'accostage. Un torseur de petits déplacements d'accostage est construit pour relier deux points $P_1$ et $P_2$ correspondants à un point $P$ de la courbe d'accostage C appartenant aux deux surfaces $S_1$ et $S_2$. Il est un torseur jeu local $\{J_{1P2}\}$ de la liaison locale en P entre les deux repères $(P_1, \vec{x_1}, \vec{y_1}, \vec{z_1})$ et $(P_2, \vec{x_2}, \vec{y_2}, \vec{z_2})$ qui coïncident au nominal. L'observateur serait alors positionné du côté $S_1$.

Le repère d'accostage est alors défini par une courbe liant les deux surfaces au nominal. Les vecteurs $\vec{y}$ et $\vec{z}$ sont définis avec les spécifications (jeu ou affleurement). Le vecteur $\vec{x}$ est alors déduit pour que le repère soit direct. On peut alors décrire mathématiquement tout type d'accostage avec déplacement (rigidifiant ou pas la courbe d'accostage C) selon qu'on veuille prendre en compte les défauts de forme de courbure des deux courbes en vis-à-vis. Dans ce cas (Figure 5.2.2), on définira un ensemble de repères d'accostages dont on pourra analyser les torseurs (type jeu) d'accostage.

Les relations entre le type d'accostage (jeu ou affleurement), le type de degré de liberté (DDL) et la dichotomie translation/rotation peut être présenté de plusieurs façons nous en proposons deux ci-après. Dans le tableau 5.2.1 nous exprimons ces six défauts en fonction avec les composantes locales du torseur des petits déplacements. Il y a trois composantes de déplacement pour chaque type d'accostage. Les jeux sont les mouvements dans le plan normal à l'affleurement et les affleurements les déplacements complémentaires.

| | Repère | Figure | TPD | Dénomination |
|---|---|---|---|---|
| **Affleurements** |  |  | $\left\{\begin{matrix} 0 & 0 \\ 0 & 0 \\ Tz & 0 \end{matrix}\right\}$ | (Dés)affleurement pur |
| |  |  | $\left\{\begin{matrix} 0 & 0 \\ 0 & Ry \\ 0 & 0 \end{matrix}\right\}$ | (Dés)affleurement évolutif |
| |  |  | $\left\{\begin{matrix} 0 & Rx \\ 0 & 0 \\ 0 & 0 \end{matrix}\right\}$ | (Dés)affleurement tangent |
| **Jeux** |  |  | $\left\{\begin{matrix} 0 & 0 \\ Ty & 0 \\ 0 & 0 \end{matrix}\right\}$ | Jeu pur |
| |  |  | $\left\{\begin{matrix} 0 & 0 \\ 0 & 0 \\ 0 & Rz \end{matrix}\right\}$ | Jeu évolutif |
| |  |  | $\left\{\begin{matrix} Tx & 0 \\ 0 & 0 \\ 0 & 0 \end{matrix}\right\}$ | Jeu tangent |

*Tableau 5.2.1 Définitions des accostages*

### 5.2.1.2   Défauts de formes des accostages

Les accostages peuvent être vus comme des déplacements de surfaces sans défauts de forme et on peut leur appliquer des méthodes d'analyse utilisant le modèle des torseurs de petits déplacements, mais si l'on veut prendre en compte les défauts de forme des contours des pièces il est nécessaire d'utiliser une modélisation prenant en compte les « déformations de contours ». Cette dernière peut être vue selon deux aspects, le point de vue comportement déformable des accostages que l'on appelle conformage et le point de vue défauts de formes des contours. Dans ce chapitre, nous nous concentrerons sur ce dernier point.

### 5.2.1.3   Application à un accostage de fenêtre de balance

Le problème posé ici est l'identification des défauts d'accostage d'un lot de fenêtres de pèse-personnes. La figure 5.2.2 montre comment on traduit un accostage en jeux et affleurements sur un ensemble de repères locaux.

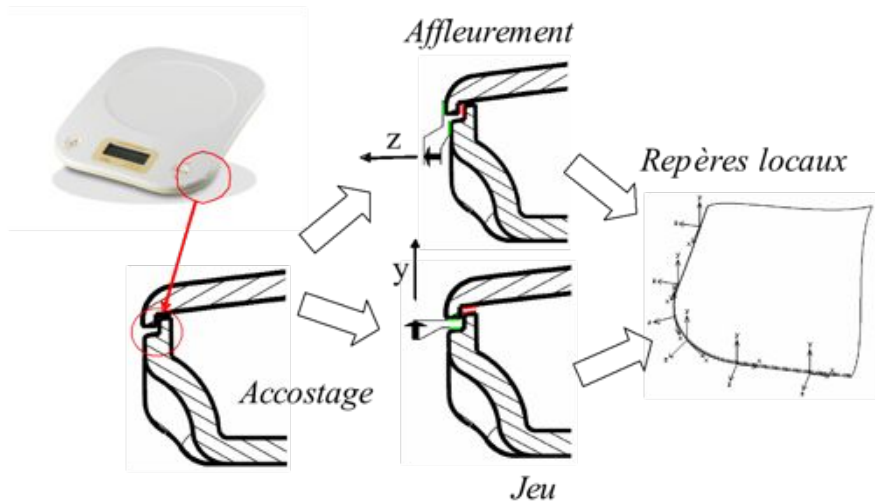

*Figure 5.2.2 Accostages d'un capot de balance de ménage*

Le modèle d'étude de cet accostage est défini à l'aide d'une poutre construite selon la courbe d'accostage. Elle est orientée selon les repères locaux. En testant (figure 5.2.3) des défauts de complexité croissante (longueurs d'ondes prépondérantes de plus en plus courtes) on a pu valider des reconstructions de qualité avec moins de cinquante modes significatifs. Ces essais ont montré l'efficience du paramétrage modal et la polyvalence de ses applications. Le terme $e$ est calculé selon.

$$e^2 = \frac{\sum_{i=1}^{n}((d_{Xmi} - d_{Xri})^2 + (d_{Ymi} - d_{Yri})^2 + (d_{Zmi} - d_{Zri})^2)}{\sum_{i=1}^{n}(d_{Xmi}^2 + d_{Ymi}^2 + d_{Zmi}^2)} \tag{5.2.1}$$

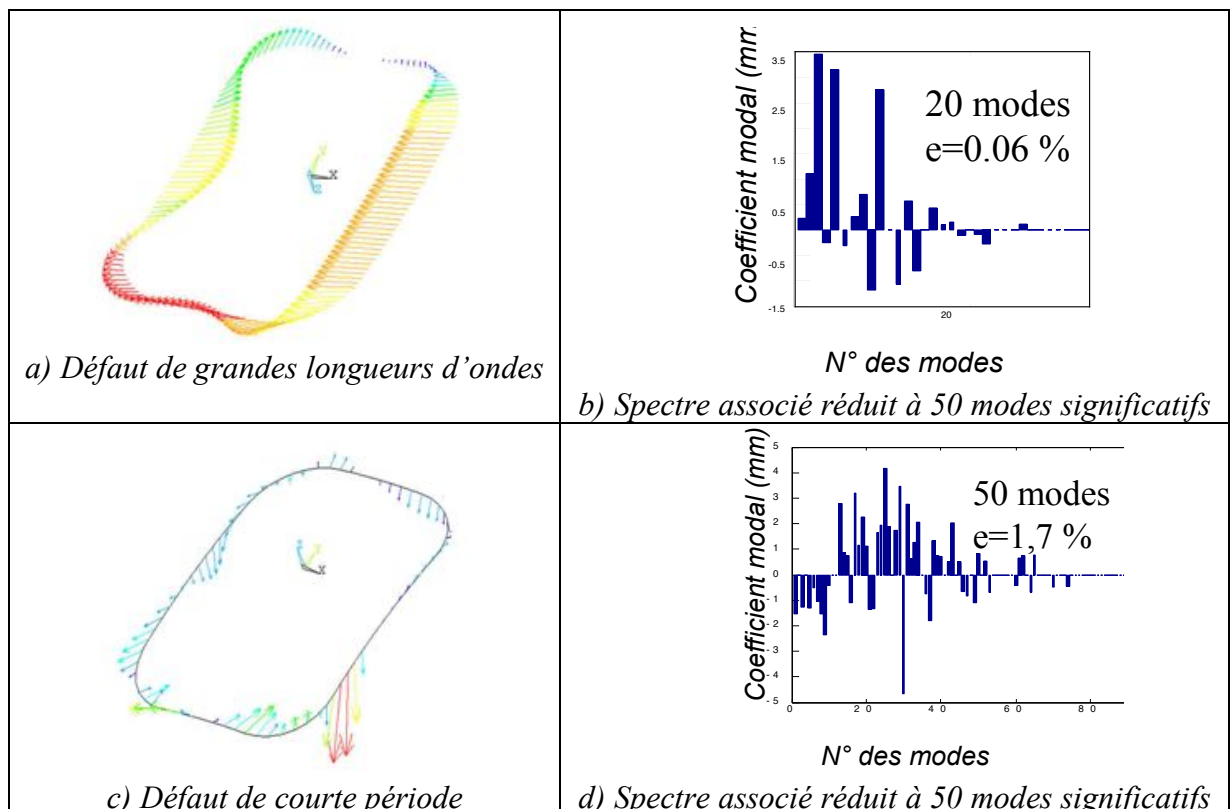

*Figure 5.2.3 Modèle poutre d'étude d'accostage*

# 6 Rugosité 3D

Les moyens de mesure ont fortement évolué [MAT 95] [MYS 03] en proposant des solutions mécaniques (microscopes AFM) et optiques (interféromètres à lumière blanche) permettant d'accéder à des micro-géométries 3D. Ces données ont fait évoluer les normes (projet de norme [ISO 25178]) qui permettront d'écrire des spécifications de rugosité 3D en prenant en compte de nouveaux critères qui enrichiront les critères 2D existants. Un nombre conséquent de scalaires y sont définis depuis la rugosité moyenne arithmétique de la surface jusqu'à des critères de direction de texture de la surface [BLA 06]. Le paramétrage modal permet d'accéder à une représentation 3D similaire à celle que nous avons développée dans l'étude des formes. En effet, si l'on désire exploiter un modèle de rugosité périodique, un paramétrage modal donnera un nombre réduit de valeurs qui peut, dans une démarche expérimentale être associé à des fonctions et des paramètres du procédé. Comme il en a été question dans le cas des ondulations, il est possible de définir un comportement anisotrope (dont la propriété des surfaces apparaît dans la future norme).

Nous ne pourrons pas, définir par la méthode modale seule, une micro-géométrie fortement anguleuse (pics et creux « pointus ») mais des méthodes basées sur les fractales [LOP 95] ou les ondelettes [LEE 98] semblent plus performantes. J'ai monté une coopération (5 séjours chercheur de 2007 à 2008 impliquant deux équipes de recherche) avec l'Université Technologique de Kaunas en Lituanie portant sur ce thème. L'étude d'un ensemble de carreaux (patches) permettra de définir les signatures moyenne et écart type de la texture 3D de la surface selon une démarche éprouvée dans l'étude d'un lot de pièces (§ 4 p.20).

# 7 Spécifications modales

La maîtrise de la qualité des formes constitue une chaîne de l'écriture des spécifications jusqu'à la métrologie. La volonté du normalisateur est de définir des outils permettant de dissocier les types de défauts. Sans aller contre cette volonté, la méthode modale peut permettre de les unifier. La zone spécifiée peut être divisée en deux zones comme le propose [AME 07], la première limite les déplacements de la surface et la seconde celle des écarts de forme. Pour aller plus loin dans cette démarche, nous pouvons alors chercher à limiter chaque forme indépendamment à la manière d'une chaîne de cotes de formes où on divise la zone globale spécifiée en autant de tronçons que de formes contributives (position, forme1, forme2, …). Nous pouvons aussi chercher à les associer pour définir un assemblage de formes dont la somme (vectorielle et non celle des zones) se partage l'espace des déplacements admissibles. Les approches présentées ci-après sont comme dans le chapitre 1, dans une démarche « au pire des cas ».

L'écriture des spécifications de formes est la partie de nos travaux qui est la plus récente. Dans [SAM 07a], nous avons écrit quelques propositions, illustrons-les avec l'exemple de la figure 5.2.1. Nous traitons l'exemple en 2D pour en simplifier les expressions.

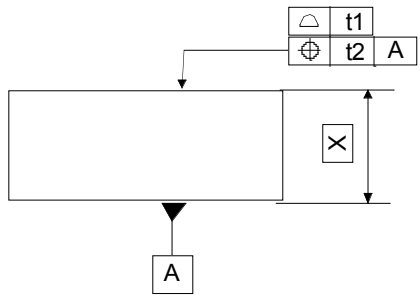

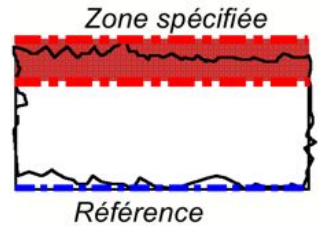

*a) Spécification*                    *b) Géométrie actuelle et zone spécifiée*

*Figure 5.2.1 Spécification et zone.*

Nous considérons une exigence fonctionnelle de localisation d'un axe dans un plan en prenant en compte les défauts de forme. Nous prenons alors une base de défauts modaux Qi, associée à un modèle poutre tel que montré dans la figure 5.2.2 (modèle poutre appuyée auquel on a ajouté les deux modes cinématiques 1 et 2).

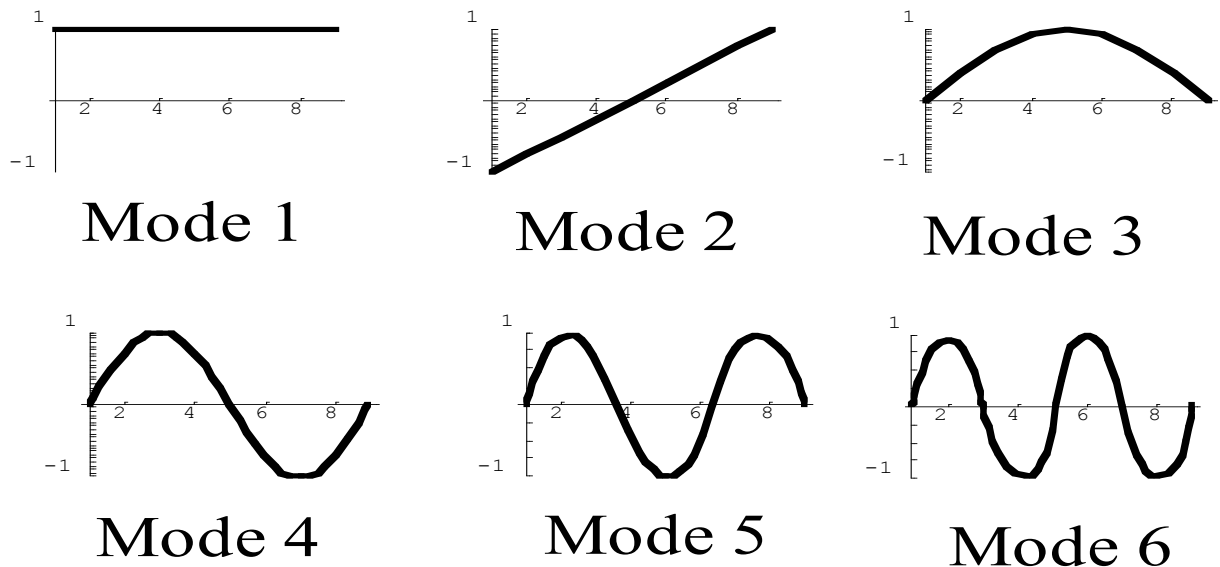

*Figure 5.2.2 Mobilités et modes d'une poutre simplement appuyée*

**Remarque :**
Les modes étant normés à l'aide de la norme infinie définie dans l'équation (2.2.4) page 10, les formes représentées ont toutes une limite égale à 1 ou (et) -1. Multipliées par la valeur de la demi-tolérance elles tangentent la zone de tolérance.

# 7.1  Tolérancement spectral

Nous pouvons écrire l'expression d'un besoin client à l'aide des coordonnées modales sous la forme d'un spectre admissible compris entre deux spectres limites. Nous avons appelé tolérancement spectral cette méthode. En analysant une forme mesurée, nous calculons les coefficients modaux et les représentons sur un spectre. La figure 7.1.1 montre trois façons de spécifier pour un spectre de coefficients modaux :
- La spécification uniforme (a) impose à chaque mode de respecter la même zone de forme en exploitant le principe métrique des coefficients modaux (ici égale à 400 µm).

- La spécification par paliers (b) permet de grouper les modes par niveaux de criticité.
- La spécification hyperbolique (c) permet de d'écrire avec une loi simple que l'exigence décroît avec l'ordre du mode.

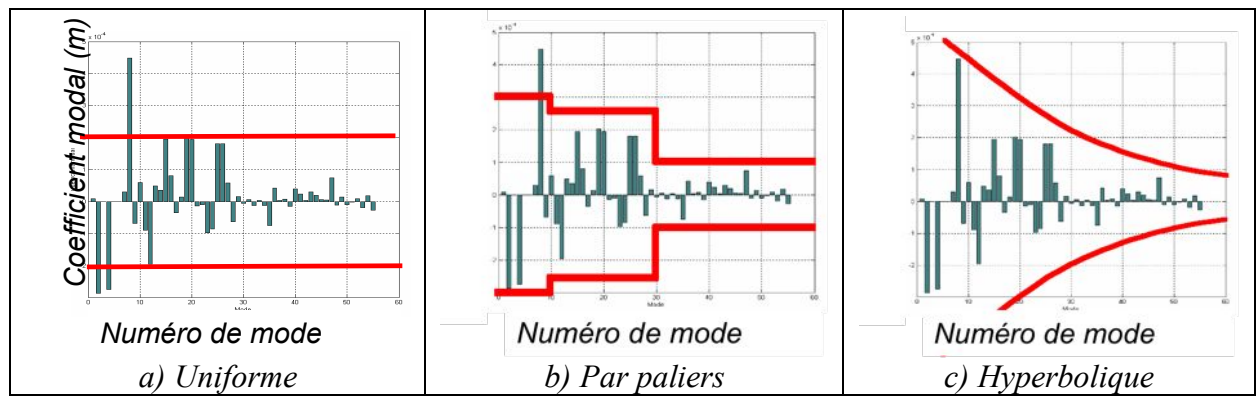

Figure 7.1.1 Spécifications spectrales.

Il va de soi que les modes rigides (associés aux TPD) peuvent être dissociés des modes de forme, pour spécifier séparément les défauts.

Cette méthode de spécification que nous pouvons nommer méthode des intervalles est simple à mettre en œuvre et très visuelle. Elle a cependant l'inconvénient de ne pas coupler les formes entre elles ce que propose la méthode des domaines modaux.

## 7.2 Domaine modal spécifié

Nous avons posé le problème de la relation entre les coefficients modaux et la limitation des défauts d'une surface à une zone. Dans ce cas il faut associer les coefficients entre eux (dont les composantes de déplacements « rigides »), en utilisant la méthode des domaines pour obtenir le domaine modal associé.

Les inéquations de la zone sont décrites par:

$$\delta_j = \sum_{i=1}^{p} \lambda_i \, Q_{ij} \qquad\qquad (7.2.1)$$

$$-\frac{t}{2} < \delta_j < +\frac{t}{2} \qquad\qquad (7.2.2)$$

Si l'on discrétise en 8 éléments (9 nœuds) la poutre modèle, et que l'on écrit les relations (7.2.2) liant les 5 premiers modes, on obtient les 18 inéquations (7.2.3) définissant le H-domaine modal associé à la spécification.

| | | Mode 1 | | Mode 2 | | Mode 3 | | Mode 4 | | Mode 5 | |
|---|---|---|---|---|---|---|---|---|---|---|---|
| Nœud 1 | -t/2 < | + 1 | $\lambda_1$ | - 1 | $\lambda_2$ | + 0 | $\lambda_3$ | + 0 | $\lambda_4$ | + 0 | $\lambda_5$ < t/2 |
| Nœud 2 | -t/2 < | + 1 | $\lambda_1$ | - 0.75 | $\lambda_2$ | + 0.38 | $\lambda_3$ | + 0.7 | $\lambda_4$ | + 0.92 | $\lambda_5$ < t/2 |
| Nœud 3 | -t/2 < | + 1 | $\lambda_1$ | - 0.5 | $\lambda_2$ | + 0.7 | $\lambda_3$ | + 1 | $\lambda_4$ | + 0.7 | $\lambda_5$ < t/2 |
| Nœud 4 | -t/2 < | + 1 | $\lambda_1$ | - 0.25 | $\lambda_2$ | + 0.92 | $\lambda_3$ | + 0.7 | $\lambda_4$ | - 0.38 | $\lambda_5$ < t/2 |
| Nœud 5 | -t/2 < | + 1 | $\lambda_1$ | + 0 | $\lambda_2$ | + 1 | $\lambda_3$ | + 0 | $\lambda_4$ | - 1 | $\lambda_5$ < t/2 |
| Nœud 6 | -t/2 < | + 1 | $\lambda_1$ | + 0.25 | $\lambda_2$ | + 0.92 | $\lambda_3$ | - 0.7 | $\lambda_4$ | - 0.38 | $\lambda_5$ < t/2 |
| Nœud 7 | -t/2 < | + 1 | $\lambda_1$ | + 0.5 | $\lambda_2$ | + 0.7 | $\lambda_3$ | - 1 | $\lambda_4$ | + 0.7 | $\lambda_5$ < t/2 |
| Nœud 8 | -t/2 < | + 1 | $\lambda_1$ | + 0.75 | $\lambda_2$ | + 0.38 | $\lambda_3$ | - 0.7 | $\lambda_4$ | + 0.92 | $\lambda_5$ < t/2 |
| Nœud 9 | -t/2 < | + 1 | $\lambda_1$ | + 1 | $\lambda_2$ | + 0 | $\lambda_3$ | + 0 | $\lambda_4$ | + 0 | $\lambda_5$ < t/2 |

(7.2.3)

Ce H-domaine est montré dans le sous espace ($\lambda_1,\lambda_2,\lambda_3$) dans la figure 7.2.1 (18 inéquations et 20 sommets 3D).

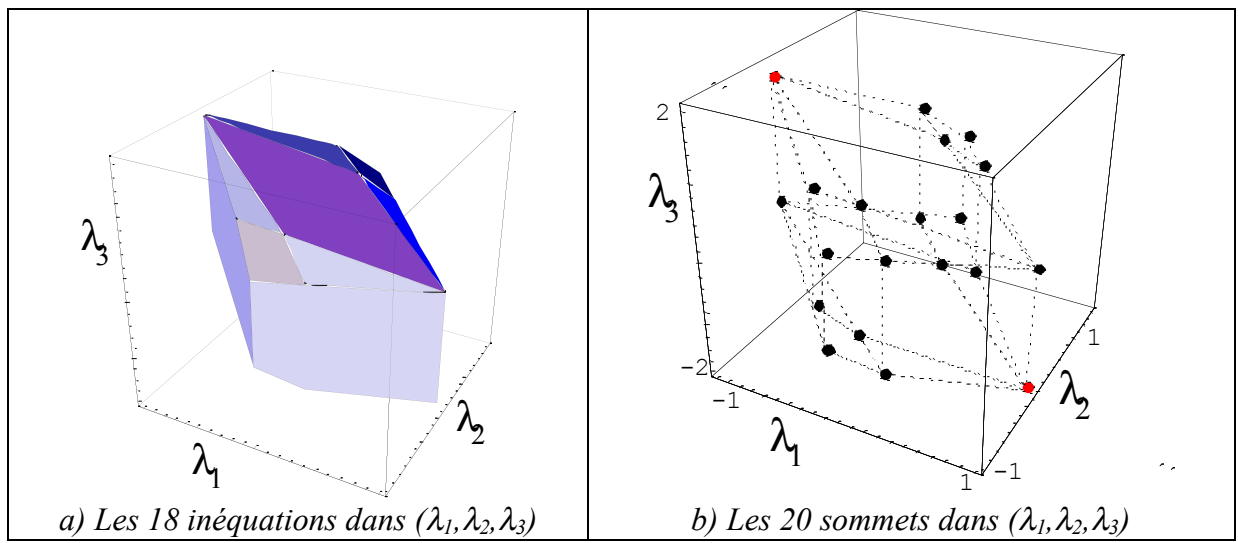

*a) Les 18 inéquations dans ($\lambda_1,\lambda_2,\lambda_3$)*      *b) Les 20 sommets dans ($\lambda_1,\lambda_2,\lambda_3$)*

Figure 7.2.1 Domaine modal spécifié

**Remarques**

- Si nous avions utilisé les six modes (travaillé en dimensions 6) de la figure 5.2.2, nous aurions eu 20 inéquations et 232 sommets.
- Chaque sommet est une composition de coefficients modaux créant une forme en contact avec la CF.
- On peut voir que $\lambda_3$ peut valoir 2 si $\lambda_1$ vaut -1 et que tous les autres $\lambda_i$ sont nuls, alors qu'il ne peut dépasser 1 (ou -1) si tous les $\lambda_i$ ($i \neq 3$) sont nuls. Cet exemple (sommets en rouge dans la figure 7.2.1.b) montre les limites du tolérancement spectral pour respecter une zone.
- Chaque sommet est une combinaison des coefficients modaux qui donne une forme en contact avec la zone spécifiée, l'ensemble des sommets est l'ensemble des combinaisons au « pire des cas ».

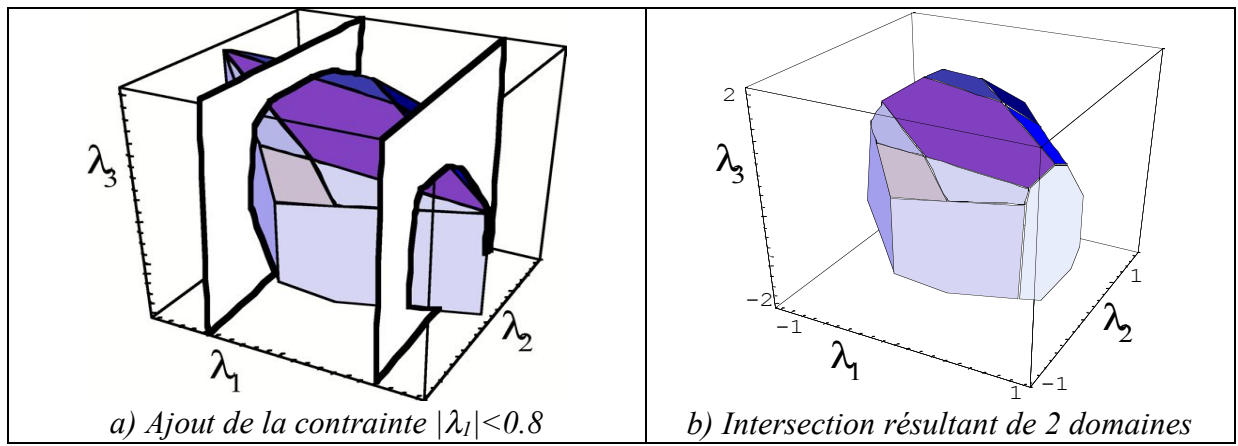

a) Ajout de la contrainte $|\lambda_1| < 0.8$                b) Intersection résultant de 2 domaines

Figure 7.2.2 Domaines spécifié et contrainte spectrale

La résolution des équations (7.2.1) et (7.2.2) appliquées à une surface discrétisée par n nœuds donnera $2_xn$ inéquations dans un espace construit par m modes propres. Le système de bi-inéquations (7.2.3) devient de taille $n_xm$ et le nombre de sommets potentiels dans un espace à m dimensions est très grand (pour un m-hypercube la formule d'Euler indique qu'il y aurait $2_xm$ faces et $4_x(m-1)$ sommets). Pour le cas d'une plaque discrétisée en $10_x10$ éléments, et l'utilisation de 20 modes (dont 3 rigides), on aurait 242 inéquations dans un espace de dimensions 20. La détermination des sommets serait longue (au plus $1,5\ 10^{17}$ sommets) et la manipulation du domaine relativement lourde.

## 7.3  Conclusion sur les spécifications modales

L'intention du concepteur peut être de limiter les écarts de formes pour des fortes périodicités et d'utiliser le tolérancement spectral qui permet d'écrire de façon simple (en limitant le nombre de paramètres à spécifier). S'il désire mixer les approches en respectant la notion de zone associée au paramétrage modal, nous proposons une extension des domaines écarts. Nous avons commencé à explorer des méthodes de simplifications automatiques des domaines qui permettront de rendre les calculs beaucoup plus rapides.
Nous savons que l'analyse au pire des cas n'est pas réaliste et son respect génère des coûts importants en sur qualité. Les approches statistiques notamment à l'aide du tolérancement inertiel proposé par Pillet [PIL 04][PIL 05] montrent une voie intéressante que nous avons commencé à explorer [ADR 07b] dans la thèse de Pierre-Antoine Adragna. Ces travaux devront être poursuivis dans le cadre des spécifications de formes en accord avec une maîtrise du Taux de Non-conformité. D'autre part l'écriture de spécifications en tolérancement modal demande à être confrontée à des exemples concrets, c'est l'objet de la troisième année (2008-2009) de la thèse d'Hugues Favrelière dans laquelle nous avons choisi de déployer la maîtrise de la qualité des formes de la métrologie vers la maîtrise statistique des procédés [FAV 07] [FAV 07a] puis vers les spécifications.

# 8  Paramétrage modal et simulations des défauts

Lors d'une analyse de structures par la MEF, on utilise la géométrie idéale. L'influence des défauts géométriques sur les résultats de l'analyse est étudiée grâce au paramétrage. Un des problèmes est de générer des défauts qui soient plausibles et non seulement faciles à générer. Nous apportons une solution qui est née dans cet environnement. Les paramétrages sont en principe liés à la construction du modèle géométrique. Ils peuvent être insuffisants lorsqu'ils n'expriment pas directement les défauts à contrôler.

## 8.1  Retour élastique en emboutissage

**Problème posé**
Les expérimentations physiques et numériques utilisent des paramètres géométriques et des paramètres de procédé. Leur mise au point se fait à travers une démarche de plans d'expériences ou de surface de réponse. Une difficulté est de construire une un paramétrage géométrique adapté au comportement du système (réglage à priori des sensibilités).
Dans [LED 06], le paramétrage modal sert à définir les formes possibles d'une pièce emboutie. Une démarche de plans d'expériences numériques (simulations par éléments finis d'opérations d'emboutissages) utilisant ce paramétrage permet de réduire considérablement les retours élastiques en définissant des outillages optimaux. La figure 8.1.1 montre le procédé (a) ainsi que la forme obtenue sans compensation du retour élastique (b).

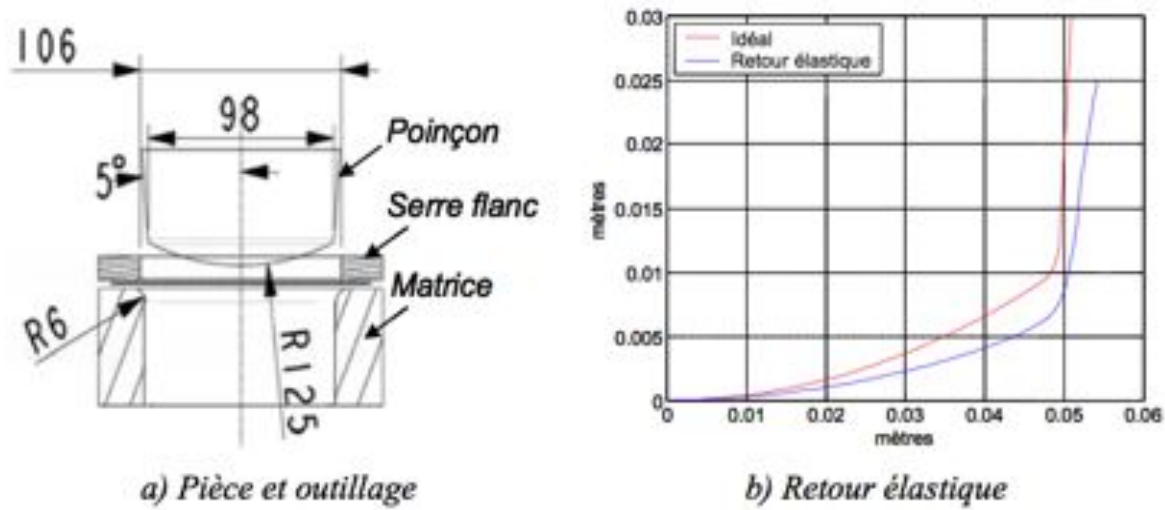

*Figure 8.1.1 Retour élastique d'une pièce emboutie*

**Démarche**
Le paramétrage initial (A,R,D,H) associé à ce procédé (Figure 8.1.2.a) est introduit dans un plan d'expériences numériques en vue de réduire le retour élastique. Le choix des paramètres est délicat (il y en a une très grande variété) et ne permet pas de réduire suffisamment le défaut (non prise en compte des formes « fléchies ») c'est pourquoi nous avons utilisé, dans une seconde approche, le paramétrage modal dont les trois premiers modes sont montrés dans figure 8.1.2.b. Il a donné de meilleurs résultats et permet en un minimum de simulations (11 après optimisation) de définir un outillage réduisant au minimum (de 6,1 mm à 0,5 mm) le retour élastique sur toute la forme.

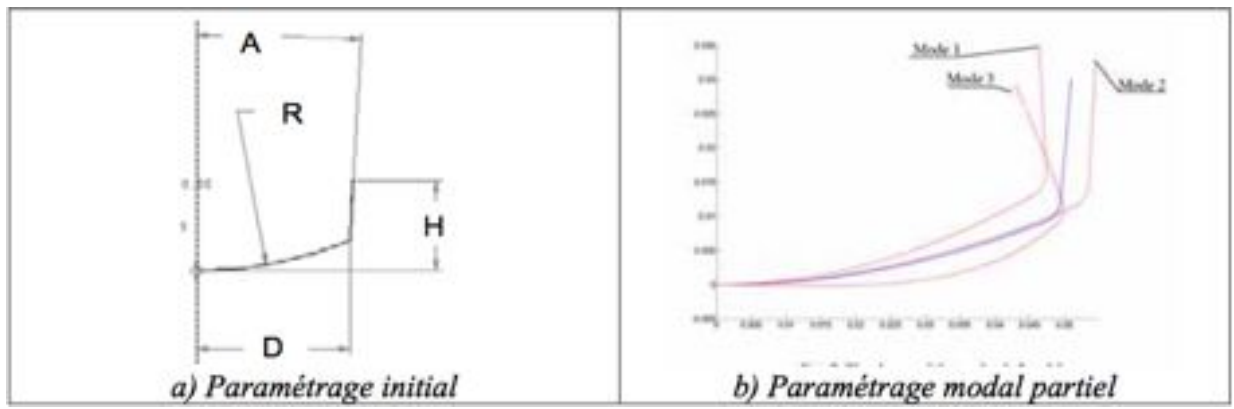

*Figure 8.1.2 Retour élastique d'une pièce emboutie*

## 8.2  Lèvres d'anisotropie en emboutissage

En emboutissant un flanc issu d'une tôle anisotrope en vue de réaliser un cylindre avec fond, il se forme des lèvres sur les bords. Dans [LED 07] , nous montrons comment le paramétrage modal des défauts de forme d'une pièce emboutie permet de limiter la taille d'un défaut issu de l'anisotropie du matériau.

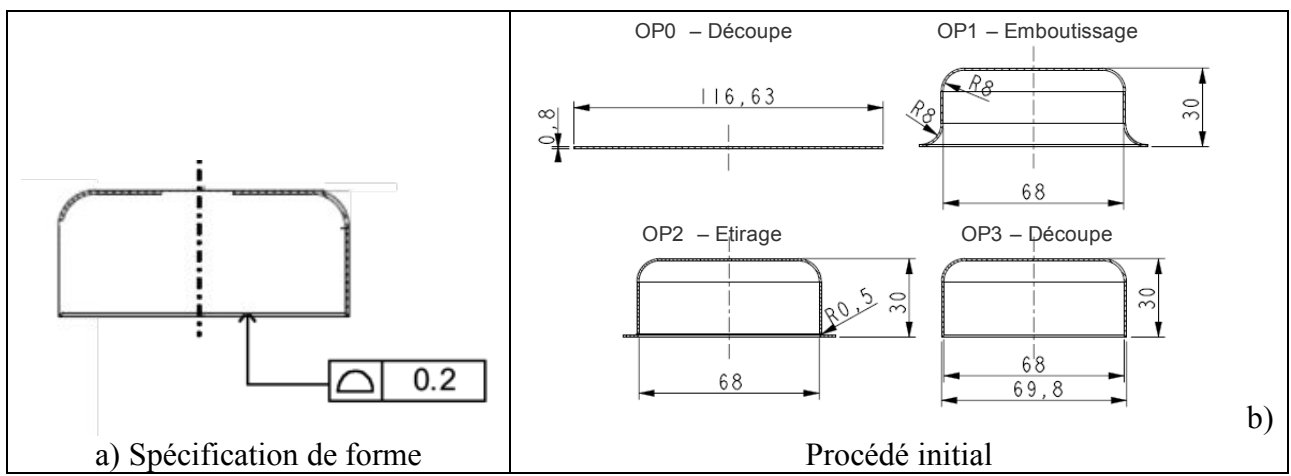

*Figure 8.2.1 Pièce emboutie*

La figure 8.2.2 montre le paramétrage du flanc (avec deux symétries) et les cinq premiers modes de la lèvre d'une pièce emboutie simulée.

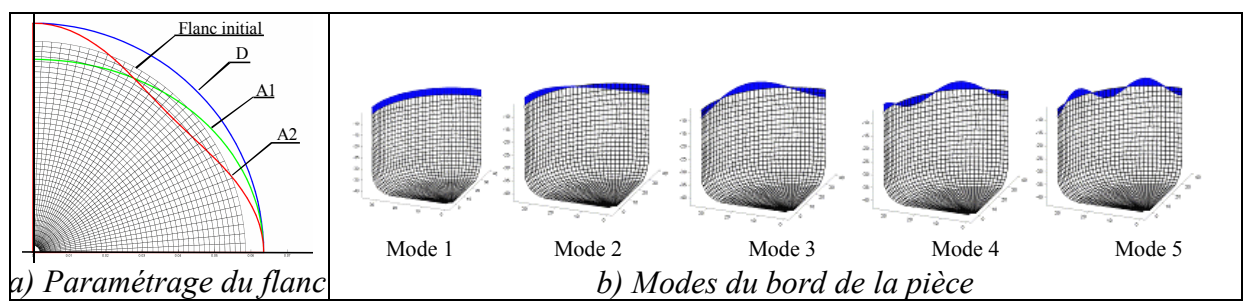

*Figure 8.2.2 Paramétrage modal du bord d'une pièce emboutie*

Grâce à l'utilisation de ce paramétrage dans une démarche de plan d'expérience, il a été possible de supprimer l'opération OP03 de découpe du bord tout en respectant la spécification de forme. Cette expérience nous laisse entrevoir des perspectives de coopération avec les spécialistes de procédés de mise en forme et leurs applications industrielles. Le paramétrage des formes peut venir en complément (ou en substitution) d'un paramétrage initial.

# 9   Le paramétrage biométrique

Dans [SAM 06], nous avons montré comment le paramétrage géométrique est ouvert au domaine biométrique. Une expression du visage est une différence (ici, soustraction de la forme «référence=neutre» à la forme «sourire») entre deux formes (Figure 8.2.1.c). La mesure de la forme est effectuée à l'aide du procédé de stéréovision mis au point par Pierre Vacher, Professeur à SYMME et Terence Coudert.

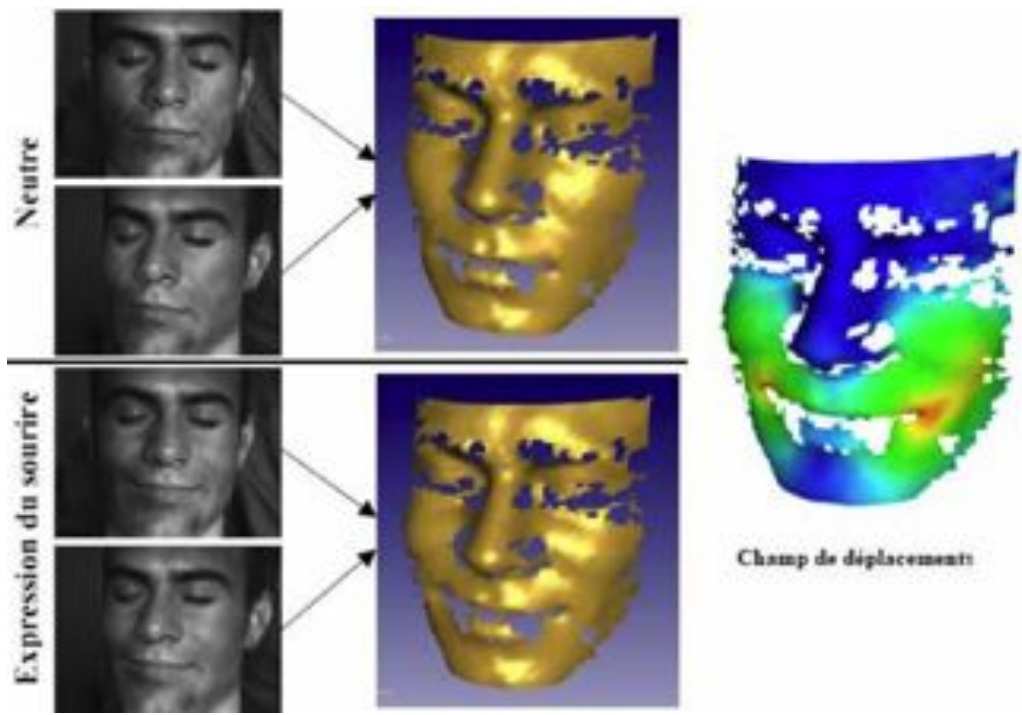

*Figure 8.2.1 Mesure d'une expression du visage*

Ce champ de déplacements peut être décomposé dans la base modale de la géométrie de référence qu'est le visage neutre (Figure 8.2.3).

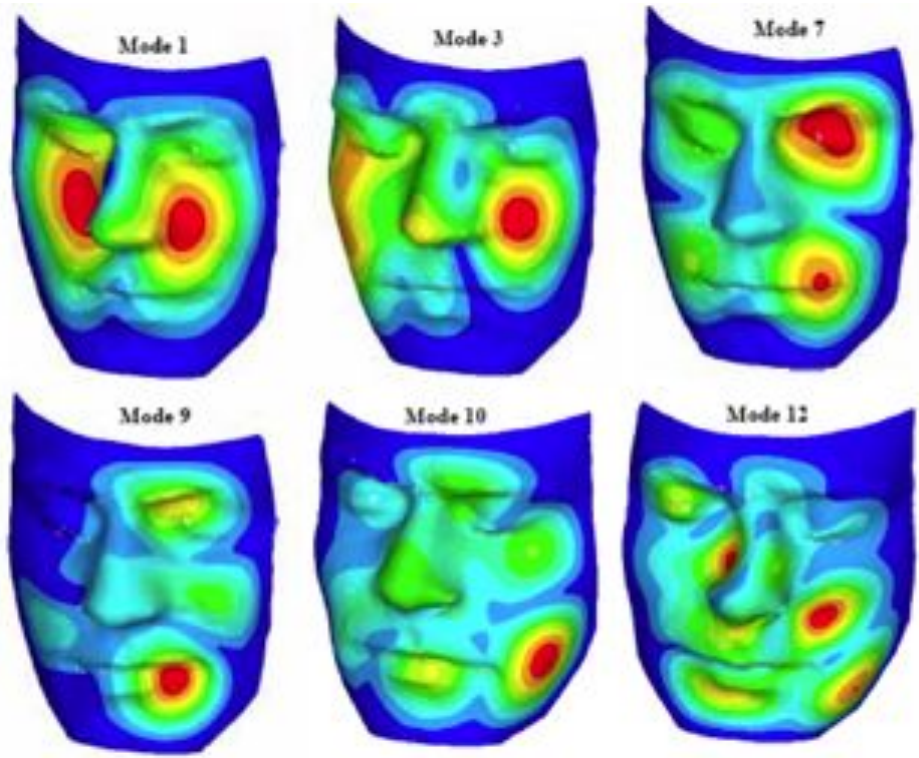

*Figure 8.2.2 Quelques modes du visage neutre*

L'expression peut alors être codée sous la forme d'un spectre (Figure 8.2.3) et être reconstruite par filtrage modal.

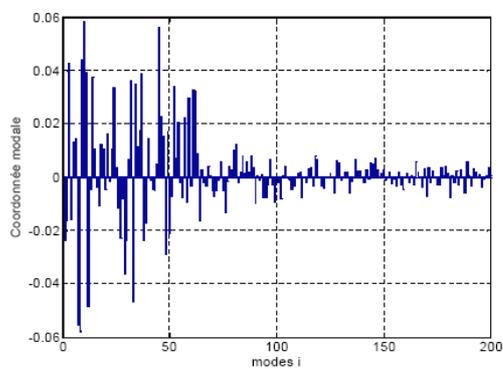
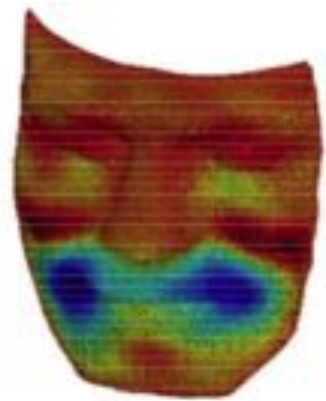

*a) Spectre du sourire*                          *b) Sourire reconstruit*

*Figure 8.2.3 Le sourire d'un visage et son spectre*

Dans la perspective de travailler sur des projets à composantes biologiques, cette étude a permis de défricher les principaux problèmes associés.

# 10 Vers le tolérancement des systèmes mécatroniques

Les assemblages se complexifient, et les outils d'aide à la conception qui ont vu le jour au cours des années 90 se diffusent dans les écoles et les industries. Le problème est toujours de définir un produit physique ayant des comportements acceptables. Le tolérancement des systèmes mécaniques peut alors se déployer vers ces systèmes.

Le groupe de travail « Modélisation Géométrique et Intégration Technologique » du Pôle Productique Rhône Alpes auquel nous avons participé avait tracé les grandes lignes d'une conception multi-vues d'un système complexe à travers l'exemple d'un moteur électrique. Ce travail a montré combien l'échange des données entre les modèles est fondamental. Nous avons proposé dans [SAM 03d] une approche où la cinématique n'est plus seulement mécanique mais où les organes de commandes apportent des variations géométriques. L'enjeu est toujours de maîtriser la géométrie mais avec des données qui ne sont plus seulement « mécaniques ». Les spécialistes en contrôle-commande peuvent corriger, moyennant un investissement en modélisation de systèmes mécaniques (quelquefois déformables) l'erreur systématique. Dans [LOT 03], nous avons appliqué ces principes à un actionneur « mécatronique ». Les incertitudes doivent souvent être encadrées faute d'être maîtrisées. Des approches combinées de modélisations sont possibles comme nous l'avons montré dans [SAM 03c] dans lequel est traité l'exemple d'un rotor guidé sur un palier magnétique en introduisant les écarts sur les pièces mécaniques de l'assemblage et les jeux dynamiques issus de l'asservissement des paliers. Plusieurs analyses sont alors possibles selon les différentes configurations d'utilisations du système, l'outil d'assemblage restant le TPD des repères de surfaces associées du système. La notion de surface associée est alors étendue à des surfaces dont les conditions ne sont plus seulement de contact.

**Etude d'une cellule accélératrice**

Dans le cadre du projet européen « tolérancement des systèmes assemblés », nous avons collaboré avec le CERN [SAM 07] sur la combinaison multi-échelle (Figure 8.2.1) des défauts d'un assemblage de cellules accélératrices du futur accélérateur linéaire CLIC (**C**ompact **Li**near **C**olider). Ces défauts sont vus depuis le niveau local (forme d'une lèvre d'un quadrant d'une cellule accélératrice) jusqu'à l'assemblage d'un ensemble de cellules sur une poutre support. Cet assemblage possède les quatre niveaux (du global vers le local) suivants de modélisation :

-        la position d'une cellule par rapport à sa voisine
-        la position d'un iris dans la cellule accélératrice
-        la position d'une lèvre (quart d'iris) dans le quadrant
-        la forme d'une lèvre d'une cavité accélératrice (approche modale)

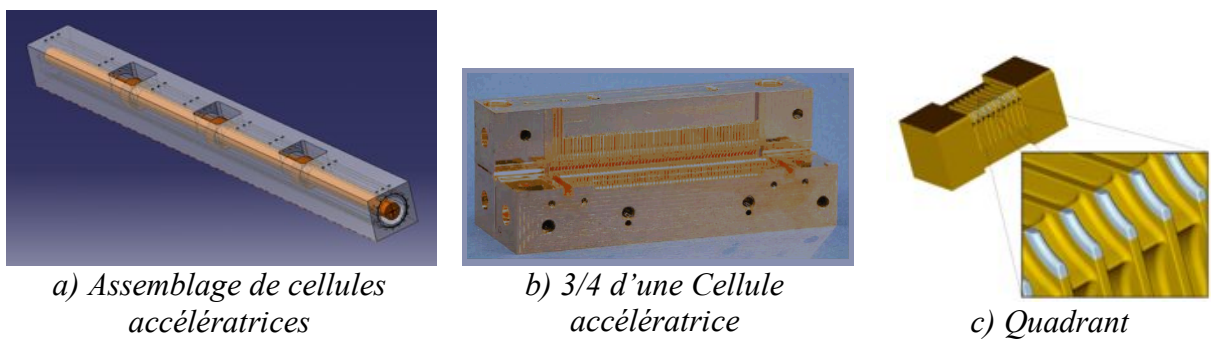

a) Assemblage de cellules          b) 3/4 d'une Cellule
accélératrices                     accélératrice                      c) Quadrant

*Figure 8.2.1 De l'assemblage au quadrant*

La qualité de l'accélération du faisceau de particules est fortement liée aux géométries. La problématique peut être qualifiée de mécatronique dans la mesure où la spécification fonctionnelle est le guidage des particules. Cette étude a eu lieu avec des données qui ont évolué durant son déroulement et qui ne sont pas stabilisées à ce jour. Il s'est posé lors de ce projet la question de la dynamique de modélisation dans un environnement multi-physique. Cette richesse laisse de nombreux points à traiter, mais un point positif est la clarification d'une méthode d'analyse de défauts en corrélation avec la définition des criticités et la description d'un langage commun (modes géométriques) aux physiciens spécialistes et aux mécaniciens du CERN. Dans la figure 3.1.3.a, un quadrant est montré dont une lèvre (isolée en traits rouges) a été paramétrée à l'aide de la méthode modale. La figure 3.1.3.b montre un mode de défaut sur une demi-lèvre de quadrant.

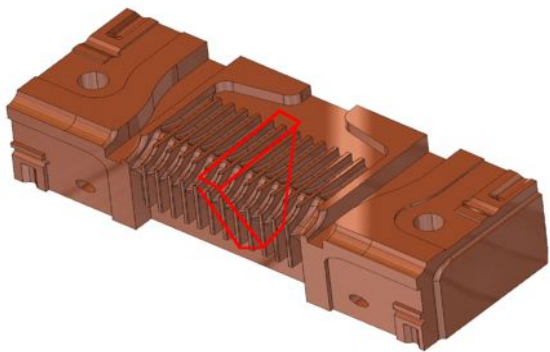
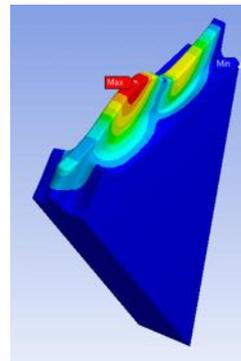

a) Quadrant et lèvre                    b) Un mode d'une demi lèvre de quadrant

Figure 8.2.2 Description modale de la géométrie locale

Il ne faut pas oublier que les systèmes mécatroniques sont bien souvent des systèmes dynamiques et les précisions sont dans ce contexte. Les approches que nous avons proposées sont des extensions des méthodes d'analyse des tolérances des systèmes mécaniques et il reste du chemin à parcourir en études interdisciplinaires pour que les acteurs d'un produit mécatronique aient à leur disposition un outil industriel de tolérancement mécatronique. L'écriture normative pourra évoluer en proposant, comme cela a été entamé pour les pièces plastiques, l'usage de labels spécifiant les conditions de mesures des assemblages.

Une voie intéressante est l'utilisation des logiciels comme solveur et de mettre en place un pré-processeur de tolérancement associé à un post-processeur. Ces deux outils restent à développer, associés à des logiciels de Conception MécaTronique Assistée par Ordinateur, dont des premières ébauches existent sous la forme d'un générateur de modèle linéaire entrées/sorties dans certains logiciels de « CMTAO ». Nous n'avons pas étudié de systèmes en quantifications statistiques mais seulement en approche au pire des cas. Il est clair que des modèles d'analyses de lots permettront de mieux connaître l'adéquation entre la performance, la fiabilité et le coût des systèmes mécatronique, ce qui reste un enjeu majeur.

# Conclusion

Le paramétrage modal est un langage des formes qui élargit l'écriture des variations géométriques en incluant les défauts de positions et de dimensions. Il définit une base mathématique simple à mettre en œuvre qui peut être exploitée à un niveau simple utilisateur ou expert. Les propriétés principales de la base de défauts sont l'exhaustivité et la valeur métrique des coefficients. Il peut être utilisé sous la forme de modes naturels si les spécifications ne précisent pas les formes à contrôler ou peut être complété par des formes spécifiques (modes technologiques). Il peut contenir les paramètres positionnels (modes rigides) et être traduit en composantes de TPD (et réciproquement). De même, on peut analyser des paramètres dimensionnels mais le domaine de prédilection est l'étude des formes de faible périodicité. En ce qui concerne les formes de forte périodicité (ondulation voire rugosité) la géométrie à analyser doit être réduite de façon à conserver la performance de la méthode (nombre réduit de modes représentant la forme).

Ce thème de recherche a germé à partir d'une idée de TP et d'une rencontre industrielle. Les aspects théoriques et la généralisation de la méthode ont ensuite été développés. Les méthodes de projection, de représentation et les spécifications modales ont un très grand champ d'applications. Les besoins inspirés des applications industrielles ont apporté de nouveaux développements tels que les coefficients métriques les développements statistiques et leurs représentations. Des projets sont en cours d'écriture pour appliquer ces méthodes à des domaines où la forme est la fonction. Notre exploration méthodologique nous a amenés à commencer par étudier la métrologie, nous continuerons par la maîtrise des procédés puis par l'écriture des spécifications.

Ces travaux ont fait l'objet d'un encadrement du mémoire CNAM d'Isabelle Perpoli (50% du travail consacré à ce thème), du co-encadrement doctoral de Pierre-Antoine Adragna (50% du travail) et de celui de la thèse d'Hugues Favrelière (100%). Ce thème est celui auquel je consacre le plus clair de mon temps recherche depuis quelques années. Les collaborations permettent de faire évoluer la méthode et montrent les pistes pertinentes à explorer.

# Références bibliographiques

*Nota : Quand le document correspondant est accessible en ligne, le* lien hypertexte *est actif sur le titre correspondant. De même, les numéros de pages sont liés aux références dans le texte.*

## Normes

| [REF] | | | Page |
|---|---|---|---|
| [ISO 10110-5] | *«Optique et instruments d'optique - Indications sur les dessins pour éléments et systèmes optiques – Partie 5 : Tolérances de forme de surface»*, ISO 10110, 2001 | …… | 5, 15 |
| [ISO 25178] | *« Spécification géométrique des produits (GPS) État de surface: surfacique - Partie 2: Termes, définitions et paramètres d'états de surface »* | …… | 6, 32 |
| [ISO 4288] | *« Spécification géométrique des produits (GPS) -- État de surface: Méthode du profil - Règles et procédures pour l'évaluation de l'état de surface »* | …… | 6 |
| [ISO 8785] | *« Spécification géométrique des produits (GPS) - Indication des états de surface dans la documentation technique de produits »* | …… | 5 |

## Articles

| Référence | Auteurs *«titre»* localisation | | Page |
|---|---|---|---|
| [ADR07] | P.A. Adragna, S. Samper, M. Pillet, H. Favreliere "A proposition of 3D Inertial tolerancing for considering the statistical combination of the location and orientation deviation" In International Journal of product development. (Acceptée le 17 déc. 2007) | …… | 26 |
| [ADR 05] | P-A. Adragna, M. Pillet, F. Formosa, S. Samper, *« Tolérancement inertiel et indices de capabilité dans la production d'assemblage »*, 4ème Conf. Int. Conception et Production Intégrées (CPI 2005), CD-ROM , Casablanca, Maroc, novembre 2005, 17 pages. | | |
| [ADR 06a] | P.A. Adragna, M. Pillet, S. Samper, F. Formosa, *« Inertial Tolerancing garantying a CPK indice on the final characteristic in an assembly production »* IDMME 2006 Grenoble, France, May 17-19, 2006 | | |
| [ADR 06a] | P-A. Adragna, S. Samper, M. Pillet, H. Favreliere, *« Analysis Of Shape Deviations Of Measured Geometries With A Modal Basis »* Journal of Machine Engineering : Manufacturing Accuracy Increasing Problems, optimisation, Vol. No. 1, 2006, pp. 134-143 | …… | 8, 20 |
| [ADR 06b] | P.A. Adragna, S. Samper, F. Formosa, M. Pillet *« Modal Tolerancing-* | | |